\documentclass[aps,prb,showkeys,longbibliography,superscriptaddress,reprint]{revtex4-1}

\usepackage{amsmath,amssymb,bm}
\usepackage{graphicx}
\usepackage{epstopdf}
\usepackage{bm}
\usepackage{dcolumn}
\usepackage{color}
\usepackage{natbib}

\begin{document}

\allowdisplaybreaks

\title{Superresolution through the topological shaping of sound with an acoustic vortex wave antenna}
\author{Matthew D. Guild}
\email{matthew.guild@nrl.navy.mil}
\affiliation{U.S. Naval Research Laboratory, Code 7160, Washington DC 20375, USA}
\author{Christina J. Naify}
\affiliation{Jet Propulsion Laboratory, California Institute of Technology, Pasadena CA 91109, USA}
\author{Theodore P. Martin}
\affiliation{U.S. Naval Research Laboratory, Code 7160, Washington DC 20375, USA}
\author{Charles A. Rohde}
\affiliation{U.S. Naval Research Laboratory, Code 7160, Washington DC 20375, USA}
\author{Gregory J. Orris}
\affiliation{U.S. Naval Research Laboratory, Code 7160, Washington DC 20375, USA}
\date{\today}

\begin{abstract}
In this paper,  we demonstrate far-field acoustic superresolution using shaped acoustic vortices.  Compared with previously proposed near-field methods of acoustic superresolution, in this work we describe how far-field superresolution can be obtained using an acoustic vortex wave antenna.  This is accomplished by leveraging the recent advances in optical vortices in conjunction with the topological diversity of a leaky wave antenna design.  In particular, the use of an acoustic vortex wave antenna eliminates the need for a complicated phased array consisting of multiple active elements, and enables a superresolving aperture to be achieved with a single simple acoustic source and total aperture size less than a wavelength in diameter.  A theoretical formulation is presented for the design of an acoustic vortex wave antenna with arbitrary planar arrangement, and explicit expressions are developed for the radiated acoustic pressure field.  This geometric versatility enables variously-shaped acoustic vortex patterns to be achieved, which propagate from the near-field into the far-field through an arrangement of stable integer mode vortices.  Two examples are presented and discussed in detail, illustrating the generation and transmission of an ``X" and ``Y" shape into the far-field.  Despite the total aperture size being less than a wavelength in diameter, the proposed acoustic vortex wave antenna is shown to achieve far-field superresolution with feature sizes 4-9 times smaller than the resolution limit.
\end{abstract}


\maketitle

\section{Introduction} \label{Sec:Intro}

Since the observation of a diffraction limit by Ernst Abbe over a century ago, overcoming this limit to achieve superresolution has been an ongoing topic of great interest in the scientific community.  While originally observed in optics, a similar diffraction limit occurs for acoustic waves.  Various mechanisms have been pursued over the years to sidestep this limitation, through either signal processing mechanisms or near-field imaging techniques to enhance the resolution.  One of the most promising advances in recent years has been in the use of acoustic metamaterials, which has enabled the realization of exotic macroscopic effective properties, including negative mass density, negative bulk modulus and negative phase speed \cite{Cummer2016, Ma2016}.  With effective material properties that are double-negative (superlens) or possess hyperbolic dispersion (hyperlens), acoustic metamaterials create negative refraction, enabling amplification of evanescent waves and subdiffraction-limit focusing \cite{Pendry2000, Zhang2008, Zhang2009, Zhu2011, GarciaChocano2014}.  However, such techniques require a negative refraction lens in the near-field, to either amplify the evanescent wave or convert it to a propagating wave.

Alternatively, helicoidal (vortex) waves can provide a method for creating stable  propagating vortices well into the far-field and the creation of features smaller than the resolution limit without the need for  an additional near-field  focusing aperture. While previous success has been achieved with superresolved optical microscopy\cite{Sheppard2004, Watanabe2004, Bokor2007}, this potential for superresolution using acoustic vortex waves has thus far not been realized. In this paper, superresolution is investigated using shaped acoustic vortices; by building on the recent advances in shaped optical vortices we propose an acoustic vortex wave antenna that is both topologically diverse and geometrically versatile.

A background on the relevant work relating to vortex waves and the methods for generating them are presented in Sec.~\ref{Sec:Background}.  A detailed development based on an acoustic leaky-wave antenna is presented in Sec.~\ref{Sec:2Dprism} for the two-dimensional (2D) case, and these results are expanded to circular and arbitrarily-shaped acoustic vortex wave antennas in Sec.~\ref{Sec:VortexPrism}.  Specifically, the use of an acoustic antenna with a single acoustic source eliminating the need for a phased array consisting of multiple active elements is discussed.

In addition to the propagating wave characteristics inside the acoustic antenna, a formulation for the radiated pressure field is presented in Sec.~\ref{Sec:VortexPrism} as well.  The radiated pressure is examined for a circular axisymmetric arrangement, and the near-field and far-field characteristics of the topological modes are examined.  In Sec.~\ref{Sec:ShapedVortex}, the theoretical formulation is expanded to arbitrarily shaped acoustic vortex wave antennas, and the acoustic vortices of two canonical shapes (a square and a triangle) are examined.  Results are presented showing how such structures enable the creation of arbitrary subdiffraction-limited shapes by the arrangement of stable integer mode vortices, to achieve far-field superresolution with a total aperture size less than a wavelength.

\section{Background} \label{Sec:Background}

First demonstrated in acoustics \cite{Nye1974, Berry2004}, vortex waves have found broad interest in optics, including use in communications, superresolution imaging and particle manipulation.  Optical vortex waves carry both spin angular momentum (SAM) due to the circular polarization of light and orbital angular momentum (OAM) associated with the helical phase fronts \cite{Allen1992}, the transmission of which enable a means for manipulation of particles through the transfer of angular momentum \cite{Yao2011} and improved communications through data multiplexing \cite{Wang2012, Bozinovic2013}.  The total strength of the vortices are characterized by the topological charge (mode), and is related to the total phase change along a closed path around the axis of the beam \cite{Berry2004, Dennis2009}.  

In addition to the transport of OAM, the highly localized vortices have been combined with radially polarized illumination to achieve superresolved optical microscopy \cite{Sheppard2004, Watanabe2004, Bokor2007}.  Superresolution has been proposed using the interaction of optical vortices with a metamaterial lens \cite{DAguanno2008}.  Utilization of the sharpened dark spot (null) along the beam axis has also been demonstrated as a means for enhanced edge-detection imaging as a vortex-based coronagraph \cite{Dennis2009, Foo2005, Swartzlander2008}.   More recently, there has been an interest in the ability to created shaped optical vortices, through either the spatial distribution of topological charge or diffractive optical elements.  Through the use of non-axisymmetric topological distributions, splitting of the on-axis vortex occurs, leading to an off-axis constellation of vortices which can be arranged into arbitrary shapes while maintaining the same topological charge \cite{Brasselet2013}. This has enabled the creation of arbitrarily shaped non-axisymmetric optical structures with sharp features, including corners, triangles, and multi-point stars \cite{ Hickman2010, Brasselet2013, Amaral2013, Amaral2014}.  

Acoustic vortex waves, like their optical counterpart, also carry orbital angular momentum, but do not exhibit spin angular momentum due to the scalar nature of longitudinal acoustic waves \cite{Thomas2003, Wilson2010}.  While comprising a somewhat less extensive body of work, there has also been sustained interest in acoustic vortex waves, focused on particle manipulation, acoustic communications and precision alignment of acoustic systems.  Acoustic vortex waves have previously been generated with either single-mode resonators \cite{Hefner1998}, or more often using a phased array, which enables a wide range of topological modes to be realized with a single aperture \cite{ Hong2015, Thomas2003, Wilson2010, Hefner1999, Marzo2015, Skeldon2008, VolkeSepulveda2008, Thomas2010}.  

Driven in part by the successes with optical vortices and the possibilities of precise wave interactions from afar, work in recent years has largely focused on the interaction of acoustic vortex waves with objects.  Such work includes investigations into the ability of acoustic vortex waves to excite the torsional modes of scatterers \cite{Skeldon2008, VolkeSepulveda2008, Marston2008, Mitri2011}, in addition to efforts determining the radiation pressures resulting from the interaction of acoustic vortex waves with submerged objects \cite{Hong2015, Marston2009}.   Recently, acoustic vortex waves were used to levitate and manipulate millimeter-sized particles in air, using a spherical phased acoustic array without the need for a confining structure to obtain the necessary radiation pressure \cite{Marzo2015}.  Recent advances have also been made with regards to the radiation forces of fractional topological modes \cite{Hong2015}, and the fractional topological modes which arise from finite time duration pulses with helical phase fronts \cite{Thomas2010}.

\begin{figure}[t!]
	\includegraphics[width=0.99\columnwidth, height=0.3\textheight, keepaspectratio]{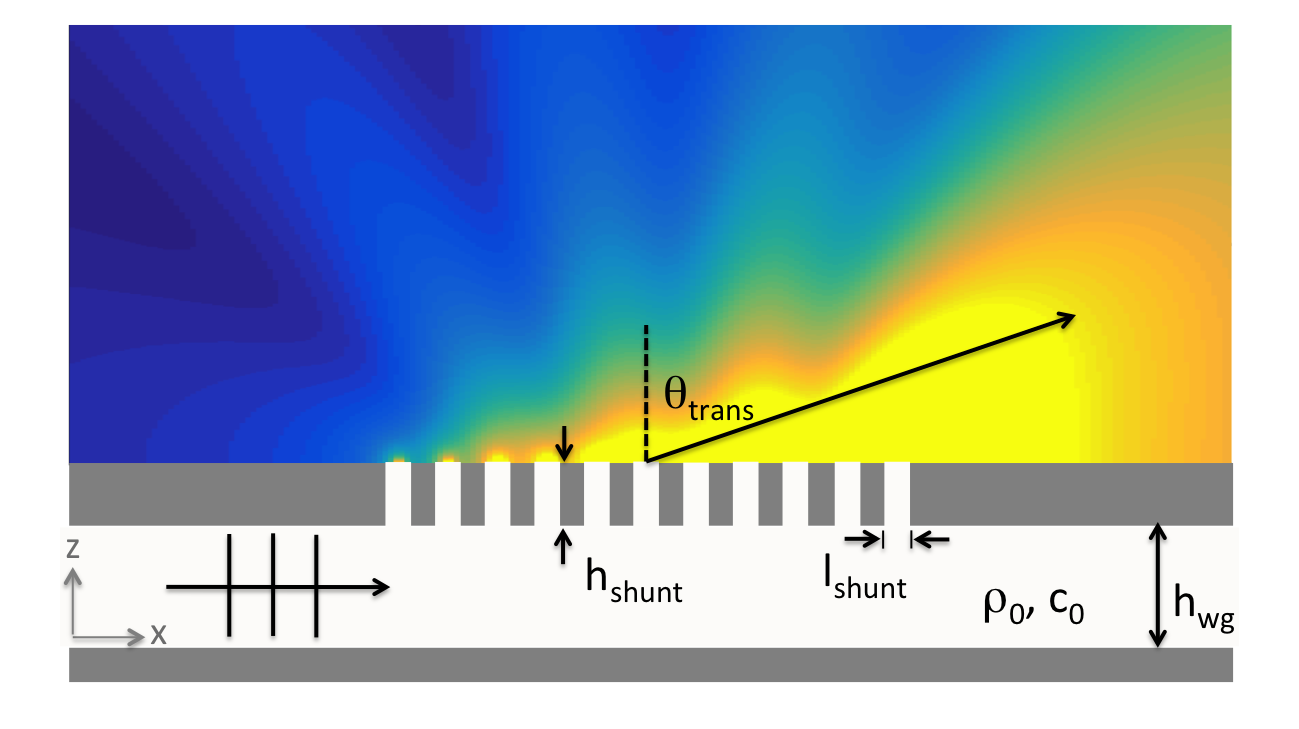}
	\caption{ (Color online) Geometry of a 2D acoustic antenna. The acoustic antenna consists of a waveguide with incident time harmonic plane wave propagating to the right (in the positive x-direction).  The acoustic waves are refracted through the arrangement of openings (acoustic shunts) into the surrounding fluid at transmission angle $\theta_{\mathrm{trans}}$.  The magnitude of a representative pressure field radiated into the surrounding fluid is shown, with the colors denoting the normalized magnitude ranging from a peak value of unity (yellow) to a minimum value of zero (dark blue).  }
	\label{Fig:Geom2D}
\end{figure}

One of the key considerations associated with vortex waves in either of acoustics or optics is the means of generating the vortex waves.  Generation of optical vortex waves are typically achieved using a spiral phase plate (SPP) or a computer-generated hologram based on diffractive optical elements \cite{Yao2011}.  Extensive work has been performed on optical vortex waves with both integer \cite{Allen1992, Yao2011, Dennis2009, Berry2005, Kotlyar2006} and fractional topological charges \cite{Berry2004, Vasnetsov1998, Lee2004, Tao2005, GarciaGracia2009, Marston2009a, Vyas2010}.  Although robust, these methods of generating optical vortex waves are limited to operation at a single mode with a fixed topological charge.  To address this issue, recent work has shown promising results for topologically diverse microwave vortex wave generation using a circular leaky-wave antenna (LWA) \cite{AlBassam2014}.  While extensively studied for microwaves \cite{Jackson2012, Monticone2015} and more recently using metamaterials \cite{Grbic2002, Liu2002, Caloz2004, Lim2004, Lai2004, Lai2006, Caloz2008, Abielmona2011, Li2011}, optical LWAs present various challenges which are the subject of ongoing research \cite{Monticone2015, Liu2010}.  In addition to electromagnetic (EM) waves, acoustic LWAs have recently been demonstrated to generate one-dimensional (1D) axisymmetric planar acoustic waves \cite{Naify2013, Esfahlani2016}, two-dimensional (2D) planar waves \cite{Naify2015} and topologically diverse circular acoustic vortex waves \cite{Naify2016}.  A detailed theoretical formulation for an acoustic antenna utilizing these recent advances with acoustic leaky-wave antennas is presented in Sec.~\ref{Sec:2Dprism}.

\section{Formulation of a two-dimensional acoustic antenna } \label{Sec:2Dprism}

An acoustic leaky wave antenna (LWA) is a device which utilizes a waveguide with a finite impedance interface, such as a series of subwavelength acoustic shunts (openings), to transmit or receive acoustic signals \cite{Naify2013, Naify2015, Naify2016,  Esfahlani2016, Bongard2010, BongardThesis}.  For the LWA, this is achieved using the refraction of the propagating signal within the waveguide into an exterior fluid medium, and therefore has been referred to as an \emph{acoustic prism} \cite{Naify2015, Esfahlani2016}.  The dispersive nature of the waveguide combined with the refraction through the interface leads to a frequency-dependent transmission angle.  A simple 2D configuration is illustrated in Fig.~\ref{Fig:Geom2D}.  The propagation of acoustic waves within the waveguide and the subsequent refraction across the interface are discussed in Sec.~\ref{Sec:2Dprop} and \ref{Sec:2Drefraction}, respectively.

\subsection{Propagation within the acoustic antenna} \label{Sec:2Dprop}
From the Helmholtz equation in cartesian coordinates, a general solution for the pressure field inside a rigid-walled waveguide is given by
\begin{equation}  \label{Eq:PressureXZ}
	P_{\mathrm{in}}(x,z,t) = p_{0} \cos(\gamma_{n} z) e^{j \left( \omega t - \beta_{x}x \right)},
\end{equation}
\noindent where $p_{0}$ is the pressure amplitude, $j \! = \! \sqrt{-1}$, $\omega$ is the angular frequency, $\beta_{x}$ is the wavenumber in the x-direction and $\gamma_{n}$ is the wavenumber in the z-direction.  From separation of variables, the relationship between the wavenumbers is
\begin{equation}  \label{Eq:SepVarBetaX}
	\beta_{x}^{2} = \left( \frac{\omega}{c_{\mathrm{ph}}} \right)^{2} = k_{0}^{2} - \gamma_{n}^{2},
\end{equation}
\noindent where $k_{0}$ is the wavenumber in the host fluid, and $c_{\mathrm{ph}}$ is the phase speed in the waveguide.  

With the wavenumber $k_{0}$ known, this means that the wavenumber $\beta_{x}$ in the direction of propagation can be evaluated once the transverse wavenumber $\gamma_{n}$ is determined.  On the interface of the acoustic antenna, the presence of the shunts will lead to a complex acoustic impedance, denoted by $Z_{\mathrm{int}}$.  For the airborne sound examined in this work, the boundary conditions at all other surfaces are assumed to be rigid.  

To determine the propagation characteristics of the waveguide, we will consider the acoustic modes in the vertical (z-axis) direction, as illustrated in Fig.~\ref{Fig:Geom2D}.  The boundary conditions for these modes are a rigid wall at the bottom surface (at $z \!=\! 0$), and an impedance condition (with complex impedance $Z_{\mathrm{int}}$) at the interface of the shunts.  Application of these boundary conditions yield a spatial dependence of the pressure in the z-direction proportional to $\cos(\gamma_{n} z)$, as given in Eq.~(\ref{Eq:PressureXZ}).  From the impedance boundary condition at $z \!=\! h_{\mathrm{wg}}$, the relationship in terms of the wavenumber $\gamma_{n}$ is given by \cite{Morse}
\begin{equation}  \label{Eq:TanGamma}
	\tan(\gamma_{n}h_{\mathrm{wg}}) = j \frac{k_{0} h_{\mathrm{wg}} Z_{0}}{\gamma_{n} h_{\mathrm{wg}} Z_{\mathrm{int}} },
\end{equation}
\noindent where $Z_{0} \!=\! \rho_{0} c_{0}$ is the acoustic impedance of the host fluid (air).  Although this transcendental equation for $\gamma_{n}h_{\mathrm{wg}}$ is indexed by the integer $n$, it is only the lowest mode ($n \!=\! 0$) which is of interest for the case of plane propagating waves, and in particular when $\gamma_{0} h_{\mathrm{wg}} \ll 1$.  Therefore, with $n \!=\! 0$ and $\tan z \approx z + (1/3)z^{3}$, Eq.~(\ref{Eq:TanGamma}) yields
\begin{equation}  \label{Eq:GammaO2}
	\left( \gamma_{0}h_{\mathrm{wg}} \right)^{2} = -\frac{3}{2} \left[ 1 - \sqrt{ 1 + \frac{4}{3} j\frac{ k_{0} h_{\mathrm{wg}} Z_{0}}{ Z_{\mathrm{int}} } } \right].
\end{equation}
When the acoustic wavelength is much larger than the height of the waveguide, Eq.~(\ref{Eq:GammaO2}) simplifies to
\begin{equation}  \label{Eq:GammaLF}
	\gamma_{0}h_{\mathrm{wg}} = \sqrt{ j\frac{ k_{0} h_{\mathrm{wg}} Z_{0}}{ Z_{\mathrm{int}} } }, \qquad k_{0} h_{\mathrm{wg}} \ll 1.
\end{equation}

Substituting Eq.~(\ref{Eq:GammaLF}) into Eq.~(\ref{Eq:SepVarBetaX}), an approximate expression for the wavenumber in the direction of propagation can be obtained such that
\begin{equation}  \label{Eq:BetaX}
	\beta_{x} \approx k_{0} \sqrt{ 1 - j\frac{1}{ k_{0} h_{\mathrm{wg}} } \frac{ Z_{0} }{ Z_{\mathrm{int}} } }.
\end{equation}
Noting that $\beta_{x} = \omega / c_{\mathrm{ph}}$, the phase speed in the waveguide can therefore be given by
\begin{equation}  \label{Eq:PhaseSpeed}
	c_{\mathrm{ph}} \approx \frac{c_{0}}{\sqrt{ 1 - j\frac{1}{ k_{0} h_{\mathrm{wg}} } \frac{ Z_{0} }{ Z_{\mathrm{int}} } } }.
\end{equation}
\noindent The phase speed is plotted as a function of normalized frequency in Fig.~\ref{Fig:Beta_v_Freq}(a).

\subsection{Refraction across the acoustic antenna interface} \label{Sec:2Drefraction}
\begin{figure}[t!]
	\includegraphics[width=0.99\columnwidth, height=0.5\textheight, keepaspectratio]{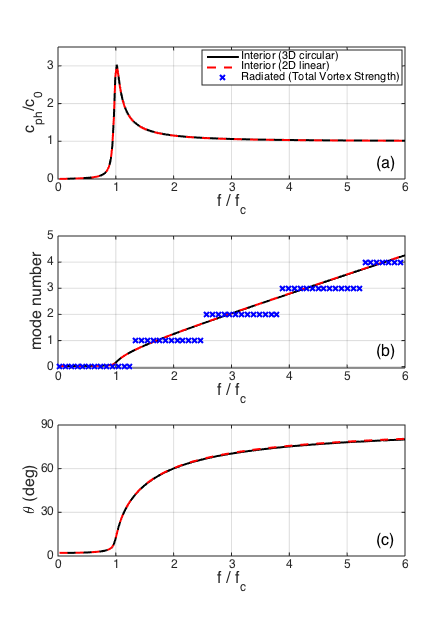}
	\caption{ (Color online) (a) Phase speed, (b) mode number and (c) transmitted angle for an acoustic antenna with the properties given in Table~\ref{Tab:Props}, plotted  as a function of the frequency normalized by the cutoff frequency given by Eq.~(\ref{Eq:CutoffFreq}).  Theoretical results are presented for interior wave propagation in a 3D circular formulation (solid line) using Eqs.~(\ref{Eq:BetaTheta})--(\ref{Eq:RadialBC}) and 2D linear formulation (dashed line) using Eqs.~(\ref{Eq:BetaX})--(\ref{Eq:ThetaRad}).  Results for the radiated field are calculated using Eq.~(\ref{Eq:VortexStrength}) for the total vortex strength.   }
	\label{Fig:Beta_v_Freq}
\end{figure}

The transmission angle $\theta_{\mathrm{trans}}$ that results from the refraction between the sound in the waveguide and the surrounding fluid can be obtained from Snell's law.  From Eq.~(\ref{Eq:PhaseSpeed}), this yields an expression for $\theta_{\mathrm{trans}}$ in terms of the acoustic antenna geometry, given by
\begin{equation}  \label{Eq:ThetaRad}
	\theta_{\mathrm{trans}} \approx \sin^{-1} \!\! \left(  \sqrt{ 1 - j\frac{1}{ k_{0} h_{\mathrm{wg}} } \frac{ Z_{0} }{ Z_{\mathrm{int}} } } \right).
\end{equation}
From Eqs.~(\ref{Eq:BetaX}) and (\ref{Eq:PhaseSpeed}), it is observed that the wavenumber and phase speed will in general be complex, leading to a wave that decays as it travels along the waveguide. In the particular case of a purely imaginary interface impedance, however, it is seen that both $\beta_{x}$ and $c_{\mathrm{ph}}$ are real, leading to a propagating wave through the waveguide and a real transmission angle according to Eq.~(\ref{Eq:ThetaRad}), which is plotted in Fig.~\ref{Fig:Beta_v_Freq}(c).

As observed in Eqs.~(\ref{Eq:PhaseSpeed}) and (\ref{Eq:ThetaRad}), the complex-valued interface impedance plays a critical role in determining the propagation through, and across the interface of, the acoustic antenna.  For the design given in Fig.~\ref{Fig:Geom2D}, the series of subwavelength shunts can be treated collectively as a distributed impedance at the surface of the interface.  The input impedance of an interface consisting of a rigid wall with shunts can be expressed as\cite{Blackstock}
\begin{equation}  \label{Eq:Zint}
	\frac{ Z_{\mathrm{int}} }{ Z_{0} } = j k_{0} \frac{ h'_{\mathrm{shunt}} }{ \phi } + \frac{ R_{\mathrm{rad}} }{ Z_{0} },
\end{equation}
\noindent where $\phi = S_{\mathrm{shunt}} / S_{\mathrm{unit}}$ is the area filling fraction of shunts in a unit cell having length $l_{\mathrm{unit}}$ and width $w_{\mathrm{unit}}$, $h'_{\mathrm{shunt}}$ is the effective height of the shunt and $R_{\mathrm{rad}}$ is the radiation impedance of the shunt.  

Note that except for the contribution of the radiation resistance, the input impedance of the shunts leads to an imaginary impedance at the surface of the interface.  Thus, assuming the effects of the radiation resistance are small, the imaginary input impedance in Eq.~(\ref{Eq:Zint}) will give a \emph{real} phase speed based on Eq.~(\ref{Eq:PhaseSpeed}), and thus lead to a propagating wave along the waveguide for a range of frequencies.  From this, it is clear that the shunts are critical in the operation and effectiveness of the acoustic antenna in air.  In addition to enabling sound to pass through the interface and thereby facilitating the transmission and reception of acoustic signals, the shunts also are a key parameter of the input interface impedance and play an important role in determining the cutoff frequency of the acoustic antenna.  Due to the difference under the square root in the denominator of Eq.~(\ref{Eq:PhaseSpeed}), a cutoff frequency will occur, below which the waves are evanescent and do not propagate through the waveguide.  From Eqs.~(\ref{Eq:PhaseSpeed}) and (\ref{Eq:Zint}), the cutoff frequency is found to be
 \begin{equation}  \label{Eq:CutoffFreq}
	f_{c} = \frac{ c_{0} }{ 2 \pi } \sqrt{ \frac{\phi}{ h_{\mathrm{wg}} h'_{\mathrm{shunt}}}  }.
\end{equation}
\noindent This cutoff frequency denotes the transition between evanescent and propagating waves within the waveguide, and is used to normalize the frequency parameter in Fig.~\ref{Fig:Beta_v_Freq}.

In addition to the impedance of the shunt, there is an additional contribution to the input impedance of the interface that arises from the acoustic waves radiated through the shunt, in the form of the radiation impedance.  The radiation impedance, $Z_{\mathrm{rad}}$, is based on the size, shape, baffle arrangement, and is in general complex: with the real part (radiation resistance, $R_{\mathrm{rad}}$) corresponding to the acoustic waves that propagate to the far-field and the imaginary part (radiation reactance, $X_{\mathrm{rad}}$) corresponding to the evanescent waves, which can be expressed as
\begin{equation}  \label{Eq:Zrad}
	Z_{\mathrm{rad}} = R_{\mathrm{rad}} + j X_{\mathrm{rad}}.
\end{equation}
Although generally complicated and often intractable to determine for arbitrary geometries, an approximate form of the radiation resistance for a long thin rectangular element with a rigid baffle is given by \cite{Mellow2011}
\begin{equation}  \label{Eq:Rrad}
	\frac{ R_{\mathrm{rad}} }{ Z_{0} } = \frac{1}{2} \frac{ k_{0} l_{\mathrm{shunt}} }{ \phi }, \qquad k_{0} l_{\mathrm{shunt}} \ll 1.
\end{equation}
It is worthwhile to note that $R_{\mathrm{rad}}$ represents the portion of the wave that is radiated from the waveguide to the far-field into the surrounding fluid.  As a result, this radiated acoustic energy represents a loss in terms of the propagating wave within the waveguide.  This can be seen in Eq.~(\ref{Eq:Zint}), with the radiation resistance contributing to the real component of the interface impedance, and therefore leading to a decay in the propagating wave along the waveguide according to Eqs.~(\ref{Eq:BetaX}) and (\ref{Eq:PhaseSpeed}).

For a source (or in this case, an external open port transmitting the acoustic wave as shown in Fig.~\ref{Fig:Geom2D}) with dimensions that are much smaller than a wavelength, the radiation impedance effects can be accounted for by using an effective height of fluid in the shunt, 
\begin{equation}  \label{Eq:Heff}
	h'_{\mathrm{shunt}} = h_{\mathrm{shunt}} + \Delta h_{\mathrm{shunt}},
\end{equation}
\noindent where $\Delta h_{\mathrm{shunt}}$ is the end correction and in general is a frequency dependent quantity resulting from the radiation reactance, $X_{\mathrm{rad}}$.  The radiation impedance will also be affected by the radiation from the 2D shape of the shunt opening.  For a single rectangular opening  (length $l_{\mathrm{shunt}}$ and width $w_{\mathrm{shunt}}$, with the width denoting the dimension in the out-of-plane direction in Fig.~\ref{Fig:Geom2D}) surrounded by a rigid baffle, the end correction for each side of the shunt is given by \cite{Ingard1953}
\begin{align} 
	 \Delta h_{\mathrm{shunt}}  = \frac{2}{\pi} \sqrt{S_{0}} & \sum_{m=0}^{\infty} \sum_{n=0}^{\infty} \nu_{mn} \left[ \frac{ \sin(m \pi \xi) }{ m \pi \xi } \frac{ \sin(n \pi \eta) }{ n \pi \eta } \right]^{2} \notag \\
	  & \times \left[  \frac{ w_{\mathrm{shunt}} }{ l_{\mathrm{unit}} } m^{2}  + \frac{ l_{\mathrm{shunt}} }{ w_{\mathrm{unit}} } n^{2} \right]^{-\frac{1}{2}}, \label{Eq:EndCorr}
\end{align}
\noindent where
\begin{equation} \label{Eq:XiEta}
\xi = \frac{ l_{\mathrm{shunt}} }{ l_{\mathrm{unit}} },  \quad \eta = \frac{ w_{\mathrm{shunt}} }{ w_{\mathrm{unit}} }, \quad S_{0} = l_{\mathrm{shunt}} w_{\mathrm{shunt}}, 
\end{equation}
\noindent and the coefficient $\nu_{mn}$ is given by $\nu_{00} \!=\! 0$ and $\nu_{0n} \!=\! \nu_{m0} \!=\! \frac{1}{2}$, with $\nu_{mn} \!=\! 1$ otherwise.

\begin{table}[b!]
	\begin{ruledtabular}
		\begin{tabular}{|ccl|}
			Parameter & Value (mm) & Description \\ \hline \hline 
			$l_{\mathrm{shunt}}$ & $2.0$ & Shunt length (tangential direction) \\ \hline 
			$h_{\mathrm{shunt}}$ & $10.0$ & Shunt height \\ \hline 
			$w_{\mathrm{shunt}}$ & $5.0$ & Shunt width (radial direction) \\ \hline 
			$h_{\mathrm{wg}}$ & $6.5$ & Waveguide height \\ \hline 
			$r_{\mathrm{in}}$ & $17.5$ & Inner radius of waveguide \\ \hline 
			$r_{\mathrm{out}}$ & $22.5$ & Outer radius of waveguide \\ \hline 
			$r_{\mathrm{mid}}$ & $20.0$ & Midline radius of waveguide \\ \hline 
			$w_{\mathrm{wg}}$ & $5.0$ & Waveguide width (radial direction) \\ \hline 
			$L$ & $125.7$ & Midline circumference ($2 \pi r_{\mathrm{mid}}$)
		\end{tabular}
	\end{ruledtabular}
	\caption{ Dimensions of the acoustic antenna examined in Sections~\ref{Sec:2Dprism}--\ref{Sec:ShapedVortex}.  In each case, the number of shunts $N$ in the acoustic antenna is 36. }
	\label{Tab:Props}
\end{table}

In Eq.~(\ref{Eq:EndCorr}), the end correction for the shunt height was given for a single shunt.  However, as described by Ingard\cite{Ingard1953}, the interaction of two (or more) elements in close proximity leads to an increase in the effective end correction.  In the low frequency limit, it was found that this converges to that of all the elements arranged side by side into one large, single element of the same area\cite{Ingard1953}.  While the exact calculations of an array of shunts are impractical to determine analytically, a reasonable approximate solution can be obtained using the results from the low-frequency limit by noting that the shunts in the acoustic antenna are deeply subwavelength.  For the acoustic antenna design under consideration in this work, the interaction of $N$ shunts corresponds to using Eqs.~(\ref{Eq:EndCorr}) and (\ref{Eq:XiEta}) with the shunt length, $l_{\mathrm{shunt}}$, and unit cell length, $l_{\mathrm{unit}}$, replaced by $N l_{\mathrm{shunt}}$ and $N l_{\mathrm{unit}}$, respectively.

\begin{figure}[t!]
	\includegraphics[width=0.99\columnwidth, height=0.3\textheight, keepaspectratio]{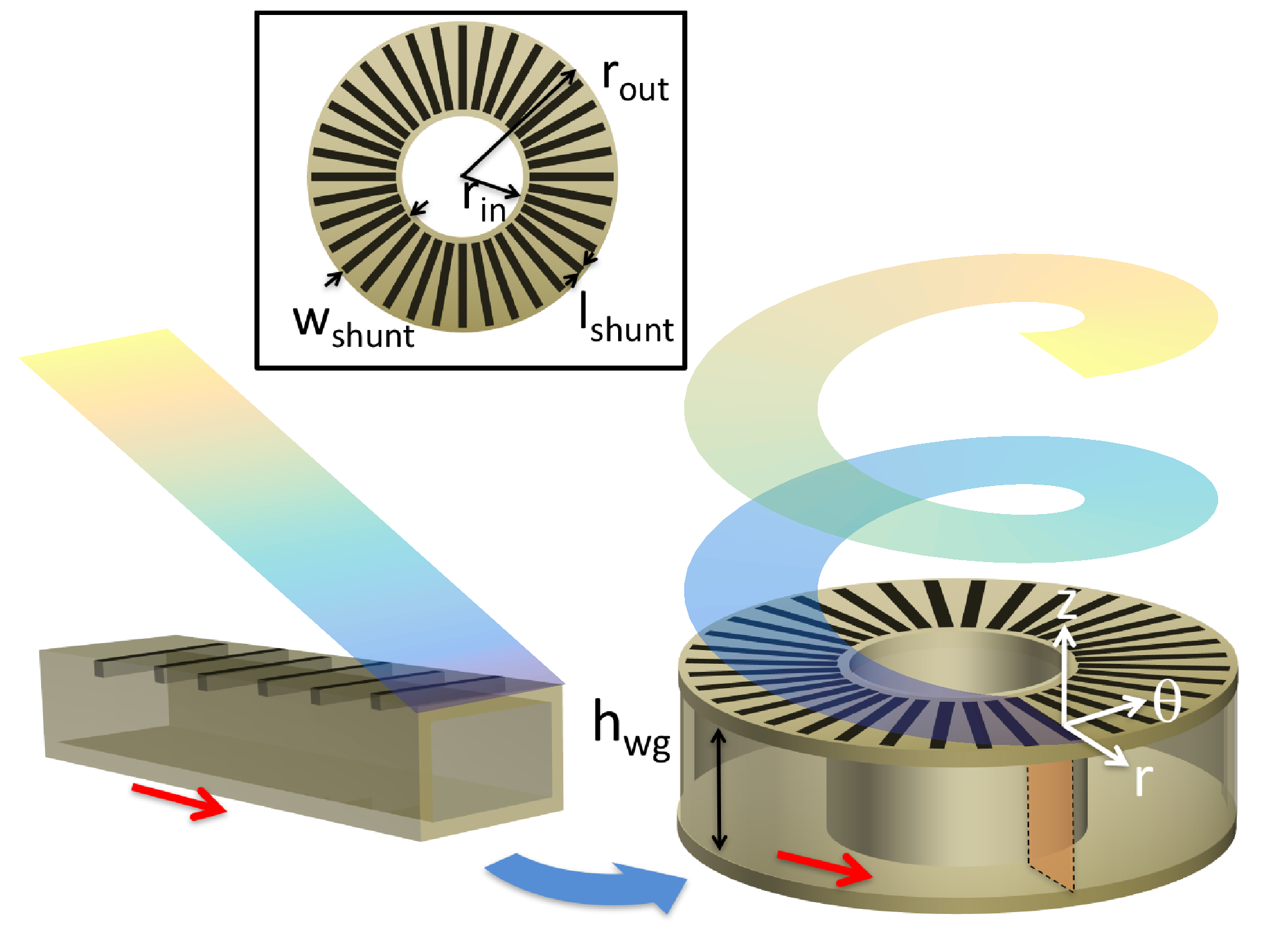}
	\caption{ (Color online) Geometry of an acoustic vortex wave antenna. By wrapping a linear acoustic leaky wave antenna (left) in a circular arrangement (right), an acoustic vortex wave can be generated.  The direction of propagation through the acoustic antenna is denoted by the red arrow.   The boundary between the inlet and outlet of the waveguide is represented by the orange-colored plane shown in the right panel.  A top view of the vortex wave antenna is illustrated in the inset. }
	\label{Fig:Geom3D}
\end{figure}

\section{Acoustic vortex wave antenna} \label{Sec:VortexPrism}

A basic acoustic vortex wave antenna configuration is illustrated in Fig.~\ref{Fig:Geom3D}.  In this case, the waveguide is curved into an annulus shape, with the sound entering into the annulus from a source, propagating around the circular path, and then exiting the annulus,  similar to the arrangement examined in Ref.~\citenum{Naify2016}.  Although the physical structure consists of circular annulus, a spacer at $\theta \!=\! 0$ prevents the interaction of sound within the waveguide between the signals entering and exiting the annulus.  This facilitates the plane progressive propagation of the sound wave along the channel, in a similar manner to that of the 2D acoustic antenna discussed in Section~\ref{Sec:2Dprism}.

\subsection{Propagation within the acoustic antenna} \label{Sec:VortexProp}

From the Helmholtz equation in cylindrical coordinates, the general solution for the pressure is
\begin{align}
	P_{\mathrm{in}}(r,\theta,z,t) = p_{0} & \cos(\gamma_{n} z) \, e^{j \left( \omega t - m \theta \right)} \notag \\
	& \times \left[  A J_{m}(k_{r} r) + B Y_{m}(k_{r} r) \right], \label{Eq:PressureRthetaZ}
\end{align}
\noindent where $J_{m}$ and $Y_{m}$ are the Bessel functions of the first and second kind, respectively, and $A$ and $B$ are coefficients determined by the boundary conditions.  Note that the wavenumber in the theta direction can be defined based on the order $m$ by
\begin{equation}  \label{Eq:BetaTheta}
	\beta_{\theta} = \frac{m}{r_{\mathrm{mid}}} = \frac{2 \pi m}{L},
\end{equation}
\noindent where $L = 2 \pi r_{\mathrm{mid}}$ is the circumference and $r_{\mathrm{mid}}$ is the radius at the midpoint of the waveguide.  Note that due to the lack of a periodic boundary condition at $\theta \!=\! 0$ and $\theta \!=\! 2 \pi$, $m$ is not restricted to being an integer value, and in general will be a non-integer and complex-valued.

The method to solve for the vertical wavenumber $\gamma_{n}$ is identical to that of the 2D acoustic antenna, and therefore Eqs.~(\ref{Eq:TanGamma})--(\ref{Eq:GammaLF}) can be used.  In a similar manner to the separation of variables in cartesian coordinates, the wavenumbers in cylindrical coordinates are related by 
\begin{equation}  \label{Eq:kr}
	k_{r}^{2} = k_{0}^{2} - \gamma_{n}^{2}.
\end{equation}
\noindent Note that from this equation, it is apparent that $k_{r}$ in cylindrical coordinates is equivalent to $\beta_{x}$ in cartesian coordinates given by Eq.~(\ref{Eq:SepVarBetaX}).  Thus, in cylindrical coordinates it is the wavenumber in the radial direction, $k_{r}$, that is equivalent to $\beta_{x}$, and the wavenumber in the theta direction $\beta_{\theta}$ (the direction of propagation) does not appear in Eq.~(\ref{Eq:kr}).  In cylindrical coordinates, the radial and angular motion within the waveguide is coupled, and can be related through the radial boundary conditions.  Assuming rigid walls at the inner and outer radii of the waveguide, the radial boundary condition can be expressed as \cite{Blackstock}
\begin{equation}  \label{Eq:RadialBC}
	J_{m}'(k_{r}r_{\mathrm{in}}) Y_{m}'(k_{r}r_{\mathrm{out}}) - Y_{m}'(k_{r}r_{\mathrm{in}}) J_{m}'(k_{r}r_{\mathrm{out}}) = 0,
\end{equation}
\noindent where the prime denotes the derivative with respect to the argument of the Bessel function.

Equation~(\ref{Eq:RadialBC}) combined with Eq.~(\ref{Eq:BetaTheta}) provides the exact relationship between $k_{r}$ and $\beta_{\theta}$.  For arbitrary waveguide dimensions these equations must be solved numerically, to implicitly determine the relationship between $\beta_{\theta}$, which appears in the order of the Bessel functions (through the mode number $m$), and $k_{r}$ which appears in the argument.  However, for most cases of practical interest, the waveguide width will be sufficiently narrow so as to support propagating waves in the theta direction without generating standing waves in the transverse (radial) direction.  

To determine an explicit approximate relationship between $k_{r}$ and $m$ (and therefore $\beta_{\theta}$), $J_{m}'$ and $Y_{m}'$ can be expanded in a Taylor series about $r = r_{\mathrm{mid}}$, in which case Eq.~(\ref{Eq:RadialBC}) can be expressed as
\begin{align}  
	&J_{m}'(k_{r}r_{\mathrm{in}}) Y_{m}'(k_{r}r_{\mathrm{out}}) - Y_{m}'(k_{r}r_{\mathrm{in}}) J_{m}'(k_{r}r_{\mathrm{out}}) \notag \\
	&\approx \varepsilon \left[ J_{m}'(k_{r}r_{\mathrm{mid}}) Y_{m}''(k_{r}r_{\mathrm{mid}}) - J_{m}''(k_{r}r_{\mathrm{mid}}) Y_{m}'(k_{r}r_{\mathrm{mid}}) \right],  \label{Eq:ApproxRadialBC}
\end{align}
\noindent where the approximation is valid for $\varepsilon^{2} \ll 1$, with $\varepsilon = k_{r}r_{\mathrm{mid}} \delta$ and 
\begin{equation}  \label{Eq:Delta}
	\delta =  1 - \frac{ r_{\mathrm{in}} }{r_{\mathrm{mid}} }  = \frac{ r_{\mathrm{out}} }{r_{\mathrm{mid}} } - 1.
\end{equation}
Noting the identity
\begin{equation}  \label{Eq:BesselCrossProd}
	J_{m}'(z) Y_{m}''(z) - J_{m}''(z) Y_{m}'(z) = \frac{2}{\pi z} \left[ 1 - \left( \frac{m}{z} \right)^{2} \right],
\end{equation}
Eqs.~(\ref{Eq:RadialBC}) and (\ref{Eq:ApproxRadialBC}) lead to the simple relationship that
\begin{equation}  \label{Eq:krBetaTheta}
	\beta_{\theta} \approx k_{r} + O(\varepsilon^{2}) \approx \beta_{x}.
\end{equation}
Therefore, to order $O(\varepsilon^{2})$, $\beta_{\theta}$ is equal to $\beta_{x}$ for a straight 2D acoustic antenna, and likewise the phase speed is given by Eq.~(\ref{Eq:PhaseSpeed}).  This result is illustrated in Fig.~\ref{Fig:Beta_v_Freq}, which shows the negligible difference between an acoustic antenna with a 3D circular waveguide (solid line) and that of a 2D linear arrangement (dashed line).

Thus, in terms of the wave propagation within the waveguide, the wavenumber is primarily determined by the vertical dimensions and shunt properties (via the complex interface impedance, $Z_{\mathrm{int}}$), and radial curvature is a secondary effect.  Although analytically investigated in this section in terms of a circular arrangement, this also holds for other geometries as well.  Even in the relatively extreme case of a right-angle bend, negligible reflections will occur for the sufficiently low frequencies at which only plane waves propagate\cite{Fahy2001}, which corresponds to the same frequency range under investigation in this work.  As a result, this enables the use of simple but powerful analytic methods to be utilized in the preliminary design of acoustic vortex wave antennas.  Note that exact values of $\beta_{\theta}$ can be obtained by numerical evaluation of Eq.~(\ref{Eq:RadialBC}) using the approximate solution from Eq.~(\ref{Eq:krBetaTheta}) as an initial guess to ensure rapid convergence.

\subsection{Radiated field of the acoustic antenna} \label{Sec:VortexRad}

A key feature of the acoustic antenna is the ability to not only control the wave propagation within the waveguide, but also effectively radiate the acoustic signals to the surrounding medium.  Previous work relating to leaky wave antennas has focused on treating them as a line array, which assumes that the waves emanate from point sources with appropriate phasing.  More advanced analysis of line arrays can also account for the diffraction from finite-sized elements using the so-called Product Theorem \cite{ShermanButler}.  While simple and robust, such an approach is restricted to the far-field of the array, which prevents analysis of the near-field Fresnel zone effects which have been predicted and observed in optics and are explored in this work.

One approach that has been extensively utilized to investigate vortex waves is the use of Laguerre-Gaussian (LG) beams.  LG beams are solutions to the paraxial wave equation, and have previously been applied to acoustic vortex waves \cite{Hefner1998,Hefner1999}.  While this accounts for both the radial and angular modes generated by the vortex wave source, these modes are restricted to integer values.  Another limiting factor is that the formulation assumes an infinite aperture, which is not appropriate for the ring-shaped acoustic antennas (with overall dimensions on the order of a wavelength or less) considered here.

Alternatively, the total pressure field radiating from an arbitrarily shaped acoustic antenna aperture can be formulated by summing the pressure radiating from each shunt.  Here we will consider a uniform rectangular source of length $l$ and width $w$, centered at the origin with a rigid baffle.  From the paraxial solution of the Rayleigh integral, an explicit expression for the radiated pressure of the element can be obtained \cite{Mast2007}
\begin{align} 
	P&_{\mathrm{elem}}(x,y,z,t) = \frac{1}{4} P_{0} \, e^{j \omega t} e^{-j \frac{1}{2} k \zeta \left[1 + \left(z/\zeta \right)^{2} \right] } \notag \\*
		\times & \left\{ \! \mathrm{erf} \! \left[ \sqrt{ j \frac{z_{l}}{\zeta} } \left( 1 \!+\! \frac{x}{l} \right) \right] + \mathrm{erf} \! \left[ \sqrt{ j \frac{z_{l}}{\zeta} } \left( 1 \!-\! \frac{x}{l} \right) \right] \right\} \notag \\*
		\times & \left\{ \! \mathrm{erf} \! \left[ \sqrt{ j \frac{z_{w}}{\zeta} } \left( 1 \!+\! \frac{y}{w} \right) \right] + \mathrm{erf} \! \left[ \sqrt{ j \frac{z_{w}}{\zeta} } \left( 1 \!-\! \frac{y}{w} \right) \right] \right\},\label{Eq:Pxyz}
\end{align}
\noindent where $P_{0}$ is the complex pressure at the face of the element, $\mathrm{erf}$ is the complex error function, $z_{l} \!=\! kl^{2}/2$ and $z_{w} \!=\! kw^{2}/2$ are the Rayleigh distances in the $x$ and $y$ directions, respectively, and
\begin{equation} \label{Eq:ZetaCases}
\zeta \!=\! 
	\begin{cases}
		\sqrt{x^{2} + y^{2} + z^{2}} \quad \text{for spherical diffraction,} \\
		\sqrt{x^{2} + z^{2}}\quad \text{for cylindrical diffraction (xz-plane).} \\
	\end{cases}
\end{equation}
The approximation of spherical diffraction gives the most accurate results for small sources, and is useful for 2D acoustic arrays of rectangular elements.  Alternatively, when one dimension of the source is small and the other is very large, such as for a 1D array, the approximation of cylindrical diffraction is the most appropriate.  This approach combines the versatility of array theory to model arbitrarily large and complicated acoustic antennas with accuracy in the near-field (Fresnel zone) obtained using paraxial solutions like those for LG beams.  Furthermore, the near-field restriction in this case is based on the size of elements, which are deeply subwavelength in scale, rather than for the size of the entire aperture, as in the solutions used in previous works on vortex waves.

To fully make use of this approach, the locations of each element must be accounted for to sum the total pressure field properly.  For the $q^{\mathrm{th}}$ element located in the source plane $z \!=\! 0$ with center ($x_{q},y_{q}$) and rotation angle $\theta_{q}$, the pressure field can be obtained by a coordinate translation and rotation to the ($\bar{x}_{q},\bar{y}_{q}$) space such that
\begin{align} 
	\bar{x}_{q} &= \ \ (x - x_{q})\cos \theta_{q} + (y - y_{q}) \sin \theta_{q}, \label{Eq:CoordTransX} \\
	\bar{y}_{q} &= -(x - x_{q})\sin \theta_{q} + (y - y_{q}) \cos \theta_{q}. \label{Eq:CoordTransY}
\end{align}

Although the overall arrangement of the shunts plays an important part of determining the total pressure field, the most important feature connecting the radiated field with the interior field of the acoustic antenna is the complex pressure term, denoted by $P_{0}$ in Eq.~(\ref{Eq:Pxyz}).  As described in Section~\ref{Sec:VortexProp}, the acoustic antenna waveguide is primarily affected by the waveguide height and shunt characteristics, with much smaller (higher order) corrections due to the curvature of the waveguide.  To obtain nearly uniform pressure across a given shunt, the waveguide should be designed with a sufficiently narrow width, $w_{\mathrm{wg}} \!=\! r_{\mathrm{out}} \!- r_{\mathrm{in}}$ (i.e. radial dimension illustrated in Fig.~\ref{Fig:Geom3D}).  

In this case, according to Eq.~(\ref{Eq:krBetaTheta}) it is therefore possible to describe the wavenumber within the curved waveguide using the 2D theory described in Sec.~\ref{Sec:2Dprism}.  Thus, from Eqs.~(\ref{Eq:PressureXZ}) and (\ref{Eq:Pxyz})--(\ref{Eq:CoordTransY}) it is possible to write an explicit approximate expression to describe the pressure radiated from an arbitrarily shaped acoustic antenna of total path length $L$ with $N$ shunts,
\begin{align}  
	P&_{\mathrm{tot}}(x,y,z,t) \approx  \notag \\*
	&   \frac{1}{4} p_{0}  \cos(\gamma_{0} h_{\mathrm{wg}}) \, e^{j \omega t} e^{-j \frac{1}{2} k \zeta \left[1 + \left(z/\zeta \right)^{2} \right] } \sum_{q=1}^{N} e^{-j \beta_{x} (q/N)L }    \notag \\*
	\times & \left\{ \! \mathrm{erf} \! \left[ \sqrt{ j \frac{z_{l}}{\bar{\zeta}_{q}} } \left( 1 \!+\! \frac{\bar{x}_{q}}{l} \right) \right] + \mathrm{erf} \! \left[ \sqrt{ j \frac{z_{l}}{\bar{\zeta}_{q}} } \left( 1 \!-\! \frac{\bar{x}_{q}}{l} \right) \right] \right\} \notag \\*
	\times & \left\{ \! \mathrm{erf} \! \left[ \sqrt{ j \frac{z_{w}}{\bar{\zeta}_{q}} } \left( 1 \!+\! \frac{\bar{y}_{q}}{w} \right) \right] + \mathrm{erf} \! \left[ \sqrt{ j \frac{z_{w}}{\bar{\zeta}_{q}} } \left( 1 \!-\! \frac{\bar{y}_{q}}{w} \right) \right] \right\}, \label{Eq:Ptot}
\end{align}
\noindent where $p_{0}$ is the source pressure in the waveguide, and
\begin{equation}  \label{Eq:ZetaBar}
	\bar{\zeta}_{q} = \sqrt{\bar{x}_{q}^{2} + \bar{y}_{q}^{2} + z^{2}},
\end{equation}
\noindent for spherical diffraction with $\bar{x}_{q}$ and $\bar{y}_{q}$ given by Eqs.~(\ref{Eq:CoordTransX}) and (\ref{Eq:CoordTransY}), respectively.  Note that approximate expressions of $\gamma_{0} h_{\mathrm{wg}}$ and $\beta_{x}$ valid for $k_{0} h_{\mathrm{wg}} \ll 1$ are given by Eqs.~(\ref{Eq:GammaLF}) and (\ref{Eq:BetaX}).

\begin{figure*}[t!]
	\includegraphics[width=0.99\textwidth, height=0.7\textheight, keepaspectratio]{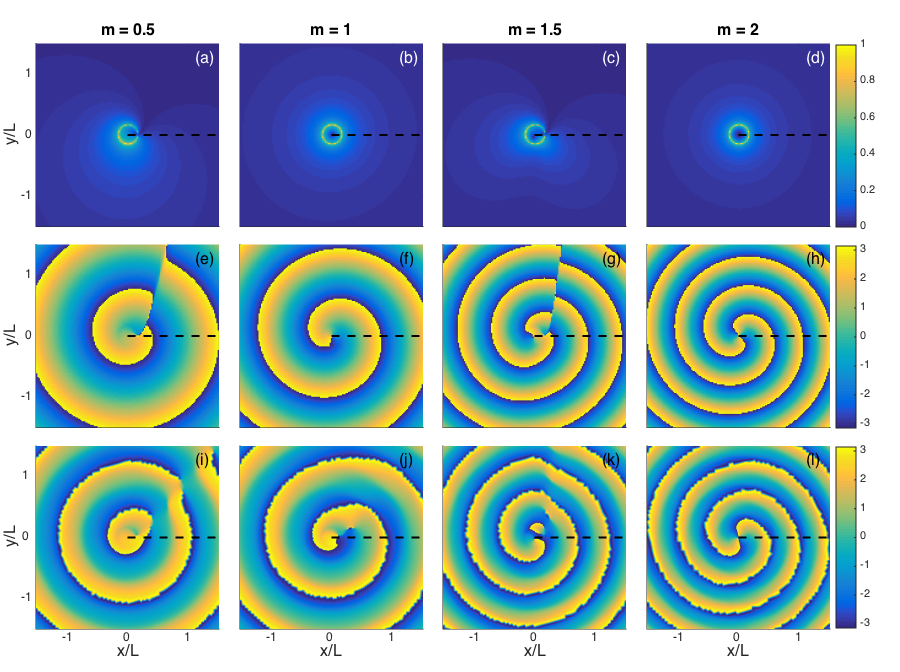}
	\caption{ (Color online) Magnitude (panels (a)--(d)) and phase (panels (e)--(l)) of the pressure field in the source plane for an acoustic vortex wave antenna in a circular arrangement at modes $m \!=\! 0.5$, $m \!=\! 1.0$, $m \!=\! 1.5$, and $m \!=\! 2.0$. Analytical results obtained using Eq.~(\ref{Eq:Ptot}) are presented in panels (a)--(h), and FEM simulations are presented in panels (i)--(l).  The $x$ and $y$ spatial coordinates of the source plane are normalized by the circumference $L$ of the acoustic antenna, and the dashed line denotes $\theta \!=\! 0$.   }
	\label{Fig:CircModes}
\end{figure*}

The implications and potential impact of Eq.~(\ref{Eq:Ptot}) are explored in the remainder of this paper.  One particularly interesting aspect of the total pressure field is phasing which occurs as the wave within the acoustic antenna propagates in a closed loop, leading to the generation of vortex waves.  When the phase term $\beta_{x}L$ does not equal an integer multiple of $2 \pi$, a phase discontinuity will occur where the start and end points meet.  The near-field and far-field vortex structures which occur under both integer and fractional modes generated by the acoustic antenna for a circular, axisymmetric shape is examined in detail in Section~\ref{Sec:VortexRad}.  An important feature regarding the use of Eq.~(\ref{Eq:Ptot}) is that it is not limited to circular shapes.  This enables the analysis of non-axisymmetric acoustic antenna arrangements, which are demonstrated to enable far-field superresolution features in Section~\ref{Sec:ShapedVortex}.

\subsection{Integer and fractional topological modes} \label{Sec:FractModes}

As demonstrated in Sec.~\ref{Sec:VortexProp}, one of the advantages of the circular acoustic antenna is its topological diverse nature, allowing one to generate a continuum of modes, including those with either integer or non-integer (fractional) values.  Even though the wave propagation within the acoustic antenna exhibits this continuum of topological modes, the radiated field of the fractional modes leads to a non-integer multiple of $2 \pi$, resulting in a phase discontinuity in the pressure.  While fractional modes have been studied in optics for many years, confusion still persists about how to properly account for these effects \cite{Marston2009a}.  Furthermore, some contradictory concepts of fractional vortices have emerged throughout the years, with theoretical results showing the total vortex strength of fractional vortices being quantized to integer values in the far-field \cite{Berry2004}, yet extensive experimental and numerical evidence indicating that optical fractional vortices lead to near-field phase discontinuities \cite{Vasnetsov1998, Lee2004, Tao2005, GarciaGracia2009, Marston2009a, Vyas2010}, including recent work expanding this to fractional acoustic vortices \cite{Hong2015}.

\begin{figure}[t!]
	\includegraphics[width=0.99\columnwidth, height=0.7\textheight, keepaspectratio]{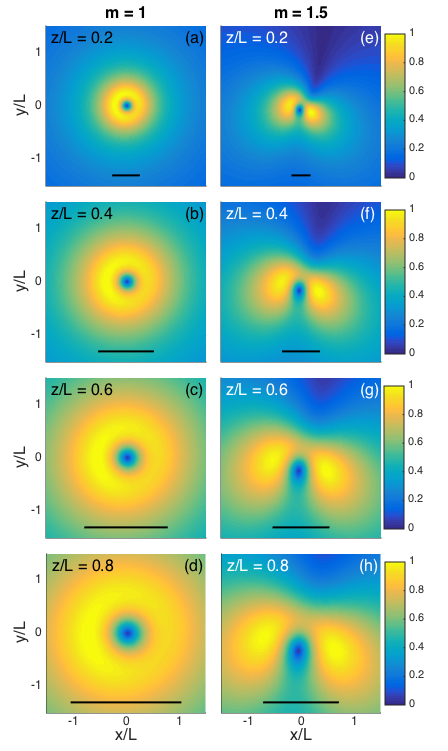}
	\caption{ (Color online) Pressure (magnitude)  obtained analytically using Eq.~(\ref{Eq:Ptot}) for an acoustic vortex wave antenna in a circular arrangement at various distances from the source at (a)--(d) $m \!=\! 1.0$, and (e)--(h) $m \!=\! 1.5$.  The distances along the $z$ direction from the source plane are normalized by the circumference $L$ of the acoustic antenna.   The corresponding Rayleigh distances from the source for each $z/L$ are tabulated in Table~\ref{Tab:RaylDist}, and the solid lines denote the resolution limit calculated using Eq.~(\ref{Eq:CircRes}). }
	\label{Fig:CircMagnitudes}
\end{figure}

Due to its ability to independently vary the frequency of operation and therefore generate a wide range of topological modes, the acoustic vortex wave antenna represents a topologically diverse aperture.  This presents an ideal means to examine the generation of both integer and non-integer topological modes.  In this section, a circular ring arrangement is used, which due to the axial symmetry, provides conceptually analogous phase features to previously examined optical vortices.  Figure~\ref{Fig:CircModes}(a)--(h) shows the magnitude and phase of the source plane at $z \!=\! 0$ for a circular acoustic vortex wave antenna based on Eq.~(\ref{Eq:Ptot}) at topological modes $m \!=\! 0.5$, $1$, $1.5$, and $2$.  In these plots, the vortices can be identified by the spiral phase fronts, with a null in the pressure magnitude at the vortex center.  For the case of integer modes, the number of ``arms" of the spiral phase fronts is equal to the mode number, and for axisymmetric arrangements, the vortices converge along the axis of the beam for integer values of $m$. 

To validate the phase variations generated in the source plane, finite element model (FEM) simulations using COMSOL multiphysics have also been performed, and are presented in Fig.~\ref{Fig:CircModes}(i)--(l).  The FEM was constructed in a similar manner to previous work by the authors\cite{Naify2016}, and consisted of a circular arrangement with the same waveguide geometry and boundary conditions as those used to obtain the analytical results.  Comparing Fig.~\ref{Fig:CircModes}(e)--(h) with Fig.~\ref{Fig:CircModes}(i)--(l), it can be seen that there is good agreement between the analytical and FEM results.  For all the topological modes ($m \!=\! 0.5$, $1$, $1.5$, and $2$), the approximate analytical results predict very similar phase fields as those determined using FEM. 

At half-step fractional modes, like $m \!=\! 0.5$ and $m \!=\! 1.5$ illustrated in Fig.~\ref{Fig:CircModes} for both the analytical results and FEM, a distinct line discontinuity is present in the phase plots.  This discontinuity can also be seen in the magnitude, which appears as a large null region in the pressure field and illustrates the formation of the next topological mode, as observed in optical vortices \cite{Berry2004}.  However, even though the discontinuity between the start and end point of the acoustic vortex wave antenna is at $\theta \!=\! 0$ (along the positive x-axis denoted by the dashed line in Fig.~\ref{Fig:CircModes}), the line discontinuity in the phase of the radiated pressure field occurs at a distinctly different angle.  Unlike the case of optical vortices formulated and generated using very large apertures such as SPPs which prescribed the discontinuity at $\theta \!=\! 0$ in the source plane, the acoustic vortex wave antenna generates the acoustic waves from a small aperture.  The resulting discontinuity in the phase is rotated in the direction of propagation within the waveguide and the resulting angle of transmission from the acoustic antenna.

\begin{figure}[t!]
	\includegraphics[width=0.99\columnwidth, height=0.7\textheight, keepaspectratio]{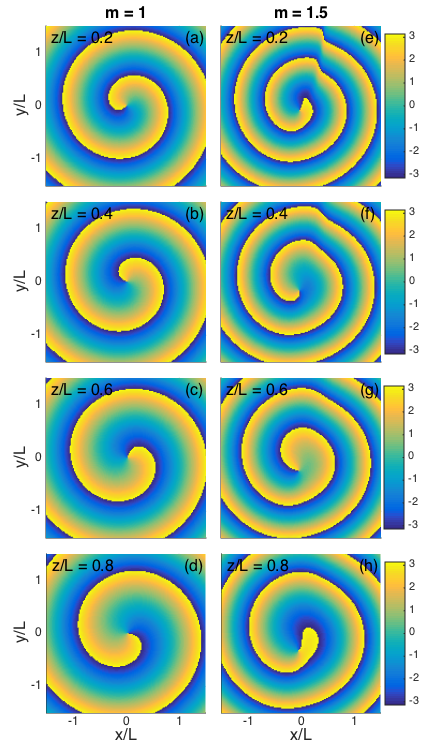}
	\caption{ (Color online) Pressure (phase)  obtained analytically using Eq.~(\ref{Eq:Ptot}) for an acoustic vortex wave antenna in a circular arrangement at various distances from the source at (a)--(d) $m \!=\! 1.0$, and (e)--(h) $m \!=\! 1.5$.  The distances along the $z$ direction from the source plane are normalized by the circumference $L$ of the acoustic antenna.   The corresponding Rayleigh distances from the source for each $z/L$ are tabulated in Table~\ref{Tab:RaylDist}. }
	\label{Fig:CircPhases}
\end{figure}

A more quantitative analysis of the topological modes can be achieved by calculation of the total vortex strength.  The total vortex strength $S_{\mathrm{tot}}$ can be determined by integrating the phase around a closed non-intersecting loop $C$, such that \cite{Berry2004, GarciaGracia2009}
\begin{equation}  \label{Eq:VortexStrength}
	S_{\mathrm{tot}} = \frac{1}{2 \pi} \oint_{C} d \varphi \frac{ \partial }{ \partial \varphi } \arg P_{\mathrm{tot}} ,
\end{equation}
\noindent where $\arg P_{\mathrm{tot}}$ is the argument of the the total radiated pressure, and $P_{\mathrm{tot}}$ can be calculated using Eq.~(\ref{Eq:Ptot}).  Previous analysis of the optical vortex strength has demonstrated that in the limit of a large closed loop (and thus far from the source), the vortex strength for an aperture of infinite spatial extent approaches a step function, resulting in a topological mode rounded to the nearest integer \cite{Berry2004}.  Thus, as the topological mode of the source is increased through a half-mode, the total radiated vortex strength far from the source jumps from one integer value to the next.

The total vortex strength plotted in Fig.~\ref{Fig:Beta_v_Freq}(b) is calculated using Eq.~(\ref{Eq:VortexStrength}) for the circular acoustic vortex wave antenna shown in Fig.~\ref{Fig:CircModes}.  From these results, it can be seen that even with a topologically diverse acoustic antenna generating both integer and non-integer modes, the radiated total vortex strength is always an integer value.  Thus, even when a phase discontinuity is present at a particular angle due to a fractional mode, the \emph{integrated} total over a closed loop will still produce a net change in phase that is an integer multiple of $2 \pi$.

\begin{table}[b!]
	\begin{ruledtabular}
		\begin{tabular}{|c|cccc|}
			 & &$z/z_{\mathrm{Rayl}}$& & \\
			$z/L$ & $m = 1$ & $m = 1.5$ & $m = 2$ & $m = 3$  \\ \hline \hline 
			$0.2$ & $2.05$ & $1.40$ & $1.18$ & $0.82$ \\ \hline 
			$0.4$ & $4.10$ & $2.80$ & $2.37$ & $1.63$ \\ \hline
			$0.6$ & $6.14$ & $4.20$ & $3.55$ & $2.45$ \\ \hline 
			$0.8$ & $8.19$ & $5.60$ & $4.73$ & $3.27$\\ \hline
			$1.0$ & $10.24$  & $7.00$ & $5.91$ & $4.08$\\ \hline
			$2.0$ & $20.48$ & $14.00$ & $11.83$ & $8.17$
		\end{tabular}
	\end{ruledtabular}
	\caption{ Corresponding values of z normalized by the Rayleigh distance for several values of $z/L$ used in this work, denoting the relative distance in the far-field.  The values for $z/z_{\mathrm{Rayl}}$ are calculated based the expression given by Eq.~(\ref{Eq:z2zR}).  }
	\label{Tab:RaylDist}
\end{table}

While the phase discontinuities from fractional modes illustrated in Fig.~\ref{Fig:CircModes} have been observed numerically and experimentally, such results have previously been limited to near-field observations of the phase.  Based on the analysis using the total vortex strength, it is apparent that the fractional vortices decay before making it to large distances away from the source.  However, there remains a disconnect in the scientific literature on this topic regarding the transition of the fractional vortices from near-field to far-field, and the connection of the discretized total source strength with the observed phase discontinuities.  

An important question, particularly related to the objective of far-field superresolution examined in Sec.~\ref{Sec:ShapedVortex}, is determining how these fractional modes decay as a function of distance from the source.  Figures~\ref{Fig:CircMagnitudes} and \ref{Fig:CircPhases} show the magnitude and phase, respectively, for a representative integer mode ($m \!=\! 1$) and fractional mode ($m \!=\! 1.5$) at different distances in the transition from the near-field to the far-field.  The distances are denoted by $z/L$, which correspond to the distance along the vertical (out of the page) $z$ direction normalized by the total aperture length (circumference) $L$, and is related to Rayleigh distance $z_{\mathrm{Rayl}}$ (denoting the distance from the source to the beginning of the far-field) in terms of the properties of the acoustic antenna according to
\begin{equation}  \label{Eq:z2zR}
	\frac{z}{ z_{\mathrm{Rayl}} } = \frac{4 \pi}{m} \frac{c_{0}}{c_{\mathrm{ph}}} \frac{z}{L},
\end{equation}
\noindent where $m$ is the mode number based on Eq.~(\ref{Eq:BetaTheta}) and $c_{\mathrm{ph}}$ and $c_{0}$ are the phase speeds in the acoustic antenna and surrounding fluid, respectively, given by Eq.~(\ref{Eq:PhaseSpeed}).  A tabulated list comparing $z/L$ and $z/z_{\mathrm{Rayl}}$ is given in Table~\ref{Tab:RaylDist} for modes $m \! = \! 1$, $m \! = \! 1.5$, $m \! = \! 2$, and $m \! = \! 3$.

In Fig.~\ref{Fig:CircMagnitudes}, negligible change is observed in the shape of the magnitude for the integer mode, except for the scaled increase in the overall dimensions due to geometric spreading.  The overall shape is similar to that of a first order LG beam, with a sharp null in the pressure field at the origin of the xy-plane due to the vortex.  Note that although the overall width of the doughnut-shaped beam is larger than the resolution limit denoted by the solid line (the details of this are discussed in Sec.~\ref{Sec:ResLimit}), the vortex null is significantly smaller than the resolution limit (about 8 times smaller than the resolution limit as tabulated in Table~\ref{Tab:Res}), and has previously been exploited in optics to achieve superresolved microscopy \cite{Sheppard2004, Watanabe2004, Bokor2007}.

\begin{figure*}[t!]
	\includegraphics[width=0.99\textwidth, height=0.7\textheight, keepaspectratio]{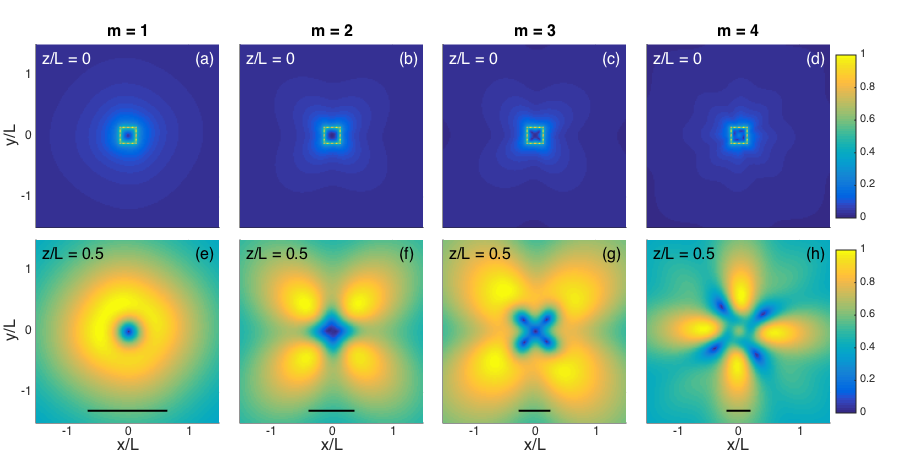}
	\caption{ (Color online) Pressure (magnitude)  obtained analytically using Eq.~(\ref{Eq:Ptot}) for an acoustic vortex wave antenna in a square-shaped arrangement at modes $m \!=\! 1$, $m \!=\! 2$, $m \!=\! 3$, and $m \!=\! 4$, in (a)--(d) the source plane, and (e)--(h) at $z/L \! = \! 0.5$.  The spatial coordinates $x$, $y$ and $z$ are normalized by the total length $L$ of the acoustic antenna, and the solid lines denote the resolution limit calculated using Eq.~(\ref{Eq:CircRes}). }
	\label{Fig:SquareModes}
\end{figure*}

This uniform doughnut-shaped ring around the vortex is not present for the case of non-integer modes.  For $m \!=\! 1.5$ as shown in Fig.~\ref{Fig:CircMagnitudes}, the peak magnitude occurs at two large lobes, which surround the sharp null of the vortex.  Note that while this null is also significantly smaller than the resolution limit, its location is shifted away from the origin due to the formation of a second vortex in the upper half of the xy-plane, resulting from the fractional component of the topological mode.  This can also be observed in the phase plots of Fig.~\ref{Fig:CircPhases}, with the progressive weakening of the line discontinuity with increasing distance from the source plane.  At even the modest change in distance from source plane (z/L = 0) to z/L = 0.8 (where $L$ is the circumference of the acoustic antenna), the fractional component has diminished significantly, and the resulting spiral phase resembles that of a single vortex with an integer vortex strength.

The results illustrated in panels (e)--(h) of Fig.~\ref{Fig:CircMagnitudes} and \ref{Fig:CircPhases} highlight the rapid decay of the fractional topological modes into the far-field.  While the total vortex strength given by Eq.~(\ref{Eq:VortexStrength}) has previously been used to emphasize that only the integer mode components propagate to the far-field, the total vortex strength of a small finite source exhibits integer values of this metric in the transition from the near-field to far-field as well.   The examination of the detailed pressure fields using Fig.~\ref{Fig:CircMagnitudes} and \ref{Fig:CircPhases} (e)--(h), however, enable a clearer picture of this change from fractional to integer modes.  As observed in Fig.~\ref{Fig:CircMagnitudes} and \ref{Fig:CircPhases} (a)--(d), the integer mode vortex remains stable and constant (excluding geometric spreading effects) in both the magnitude and phase of the pressure field throughout the transition from near-field to far-field.  In addition to their robustness, these integer modes also demonstrate vortex nulls that are much smaller (more than a factor of 8 times smaller for $m \!=\! 1$) than the resolution limit, and therefore offer the possibility of creating far-field sub-resolution limit features.

\section{Shaped acoustic vortices and superresolution} \label{Sec:ShapedVortex}

In Sec.~\ref{Sec:FractModes}, localized nulls in the pressure field generated by circular acoustic vortex wave antennas were observed and examined, and shown to be smaller than the resolution limit.  While these localized nulls have been used in optics as a means to obtain superresolution, the sub-resolution limit null is surrounded by a large doughnut-shaped peak, which is larger than the resolution limit.  As a result, such an approach is limited to creating single sub-resolution points, yet does not enable generating more complex sub-resolution structures useful for communications or particle manipulation applications.  While recent work has begun to examine the ability to manipulate the acoustic pressure field from circular arrays using fractional mode shaped acoustic vortices \cite{Hong2015}, the rapid decay of fractional mode components discussed in Sec.~\ref{Sec:FractModes} limit such an approach to near-field applications.  

Alternatively, shaped acoustic vortices can be generated using a non-axisymmetric source, which has previously been examined in optics using the spatial distribution of topological charge or diffractive optical elements\cite{ Hickman2010, Brasselet2013, Amaral2013, Amaral2014}.  Unlike these optical means, which are limited to operation at single topological modes, the acoustic antenna described in this work enables a topologically diverse method of generating the vortex waves.  Furthermore, the theoretical formulations developed in Sec.~\ref{Sec:VortexPrism} enable the design of arbitrarily shaped acoustic antenna arrangements, facilitating a geometrically versatile means to shape the acoustic vortices.  In this section, the use of shaped acoustic vortices using an acoustic vortex wave antenna is explored as a means for generating superresolved features and shapes.  This analysis begins with quantifying the metric of the resolution limit, which is examined in Sec.~\ref{Sec:ResLimit}.  In Sec.~\ref{Sec:Shaped}, the generation of shaped acoustic vortices is discussed, followed by a demonstration of superresolution from a square-shaped and triangular-shaped ring.

\begin{figure}[t!]
	\includegraphics[width=0.99\columnwidth, height=0.7\textheight, keepaspectratio]{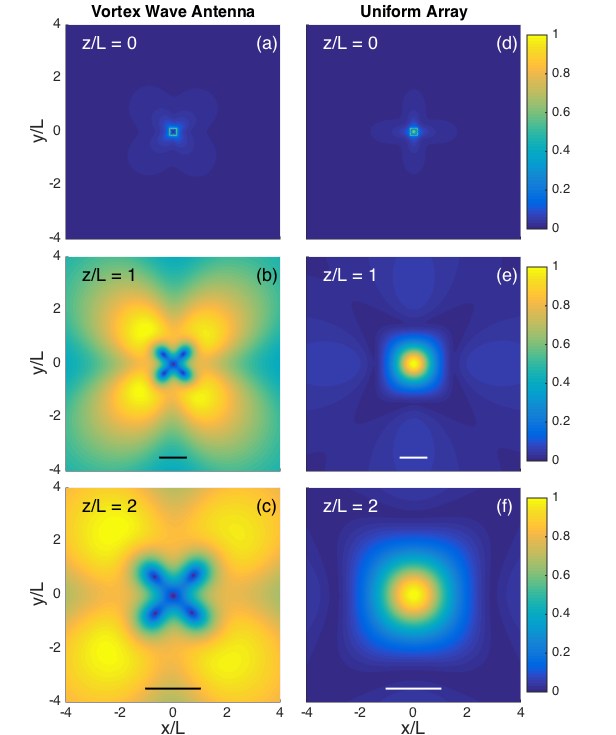}
	\caption{ (Color online) Pressure (magnitude)  obtained analytically using Eq.~(\ref{Eq:Ptot}) for a square-shaped arrangement at the mode $m \! = \! 3$ at various distances from the source for (a)--(c) an acoustic vortex wave antenna, and (d)--(f) a uniform acoustic array.  The distances along the $z$ direction from the source plane are normalized by the circumference $L$ of the acoustic antenna.   The corresponding Rayleigh distances from the source for each $z/L$ are tabulated in Table~\ref{Tab:RaylDist}, and the solid lines denote the resolution limit calculated using Eq.~(\ref{Eq:CircRes}). }
	\label{Fig:SquareMagnitudes}
\end{figure}

\subsection{Resolution limit} \label{Sec:ResLimit}

The resolution limit for an acoustic source can be obtained from the beam width of the radiated pressure field.  Due to diffraction,  a discrete finite source will not produce a perfectly collimated beam, but rather exhibits a roll-off from the on-axis peak value down to some local off-axis minimum.  Different criteria have been used over the years to denote the width of the beam, and in this work we will use the 3 dB point (half power, or where the amplitude is 0.707 the peak value) to define the edge of the beam, and therefore denotes the resolution limit of the source.  The 3 dB resolution, $d_{r}$, at a distance $z$ from the source can be obtained theoretically for an acoustic aperture of effective radius $r_{\mathrm{eff}}$, which can be expressed as \cite{Kino}
\begin{equation}  \label{Eq:CircRes}
	d_{r} = 0.51 \frac{ \lambda_{0} z}{ r_{\mathrm{eff}} } = 1.02 \pi \frac{ z}{ k_{0} r_{\mathrm{eff}} },
\end{equation}
\noindent where $\lambda_{0}$ and $k_{0}$ are the wavelength and wavenumber in the surrounding fluid, respectively.

\begin{figure}[t!]
	\includegraphics[width=0.99\columnwidth, height=0.7\textheight, keepaspectratio]{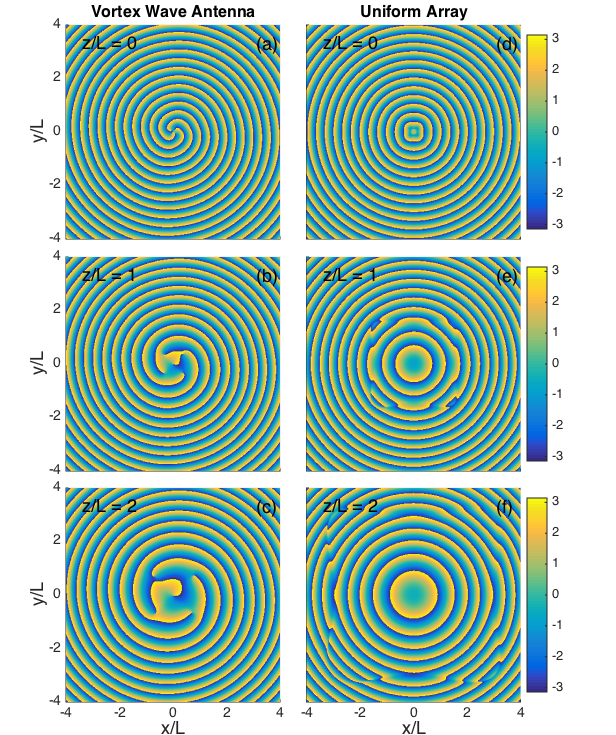}
	\caption{ (Color online) Pressure (phase)  obtained analytically using Eq.~(\ref{Eq:Ptot}) for a square-shaped arrangement at the mode $m \! = \! 3$ at various distances from the source for (a)--(c) an acoustic vortex wave antenna, and (d)--(f) a uniform acoustic array.  The distances along the $z$ direction from the source plane are normalized by the circumference $L$ of the acoustic antenna, with the corresponding Rayleigh distances from the source for each $z/L$ are tabulated in Table~\ref{Tab:RaylDist}. }
	\label{Fig:SquarePhases}
\end{figure}

The resolution limit given by Eq.~(\ref{Eq:CircRes}) applies to both unfocused (circular) and focused (spherical) acoustic sources.  Although the resolution limit is the same for focused and unfocused sources in this case, the use of a focused acoustic source allows for a significant increase in the amplitude in the focal region.  From Eq.~(\ref{Eq:CircRes}) it can be seen that the resolution can be made smaller than the diameter of the source at distances $z \! \lesssim \! 0.8 \, z_{\mathrm{Rayl}}$, where $z_{\mathrm{Rayl}} \!=\! kr_{\mathrm{eff}}^{2}/2$ is the Rayleigh distance denoting the start of the far-field.  However, this reduction in resolution compared to the diameter is limited to the near-field, and in the far-field geometric spreading will lead to an increase in $d_{r}$ with increasing distance from the source.

\begin{figure*}[t!]
	\includegraphics[width=0.99\textwidth, height=0.7\textheight, keepaspectratio]{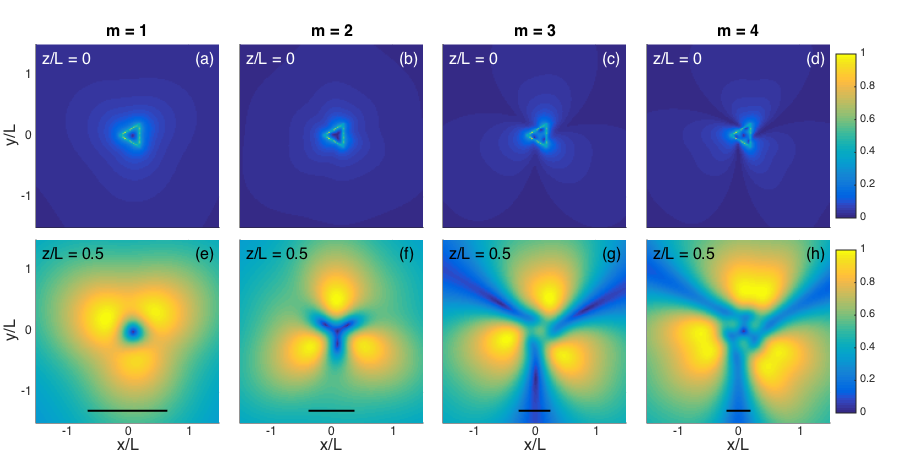}
	\caption{ (Color online) Pressure (magnitude)  obtained analytically using Eq.~(\ref{Eq:Ptot}) for an acoustic vortex wave antenna in a triangular-shaped arrangement at modes $m \!=\! 1$, $m \!=\! 2$, $m \!=\! 3$, and $m \!=\! 4$, in (a)--(d) the source plane, and (e)--(h) at $z/L \! = \! 0.5$.  The spatial coordinates $x$, $y$ and $z$ are normalized by the total length $L$ of the acoustic antenna, and the solid lines denote the resolution limit calculated using Eq.~(\ref{Eq:CircRes}). }
	\label{Fig:TriangleModes}
\end{figure*}

\subsection{Shaped acoustic vortices} \label{Sec:Shaped}

In addition to the circular, axisymmetric arrangements explored in Sec.~\ref{Sec:VortexPrism}, acoustic vortex wave antennas can be constructed into arbitrary shapes, enabling the creation of shaped acoustic vortices.  Vortices created using circular or axisymmetric sources converge along the central axis for integer topological modes, as discussed in Sec.~\ref{Sec:FractModes} and illustrated in Fig.~\ref{Fig:CircMagnitudes}.  However, shaped optical vortices have been shown to exhibit vortex splitting for non-axisymmetric arrangements at higher topological modes \cite{Brasselet2013}.  Such a feature has not previously been exploited for acoustic waves, and the use of the acoustic vortex wave antenna enables multiple topological modes to be investigated using a single aperture.

\begin{table}[b!]
	\begin{ruledtabular}
		\begin{tabular}{|cccc|}
			Aperture & Mode & Shape & Resolution \\ \hline \hline 
			Circular ring & $1$ & point & $0.12 \, d_{r}$ \\ \hline 
			Square ring & $1$ & point & $ 0.11 \, d_{r}$ \\ \hline
			Square ring & $3$ & ``X" & $0.23 \, d_{r}$ \\ \hline 
			Triangular ring & $1$ & point & $ 0.12 \, d_{r}$ \\ \hline
			Triangular ring & $2$ & ``Y" & $0.19 \, d_{r}$
		\end{tabular}
	\end{ruledtabular}
	\caption{ Resolution, mode number and resulting far-field pressure shape for the different acoustic antenna arrangements examined in this work.  The resolution is determined using a 3-dB criteria from the pressure magnitudes illustrated in Fig.~\ref{Fig:CircMagnitudes}, \ref{Fig:SquareModes}, \ref{Fig:SquareMagnitudes}, \ref{Fig:TriangleModes}, and \ref{Fig:TriangleMagnitudes}, and is written as a function of the resolution limit, $d_{r}$, for a traditional source given by Eq.~(\ref{Eq:CircRes}).   }
	\label{Tab:Res}
\end{table}

In general, the design and numerical evaluation of such acoustic pressure fields would be computationally quite time consuming.  For each iterative design, the 3D acoustic pressure field would need to calculated over a wide range of frequencies to determine the modal frequencies of interest.  For the case of a topologically diverse aperture such as the acoustic vortex wave antenna, the numerical meshing of the source would need to be sufficiently small to capture the detail of each radiating element.  At the same time, the range of geometric parameters that govern the dispersive wave propagation within the acoustic antenna, including those listed in Table~\ref{Tab:Props}, would each need to be varied and optimized to achieve a desired performance.  While technically possible, such a numerical approach would prove impractical to explore the design space of complex arrangements.  Alternatively, the theoretical formulation developed in Sec.~\ref{Sec:VortexPrism} enables one to determine the wavenumber within the acoustic antenna, and from this determine the corresponding magnitude and phase at each radiating shunt.  For a given arbitrary arrangement of the acoustic antenna, the total radiated pressure (valid in the near-field and far-field of the aperture) can then be explicitly calculated from Eq.~(\ref{Eq:Ptot}), as was done for the circular arrangement examined in Sec.~\ref{Sec:FractModes}.

Throughout the remainder of this section, two representative cases will be examined to demonstrate the effectiveness of this approach and capabilities for achieving far-field superresolution using shaped acoustic vortices: (1) a square-shaped ring and (2) a triangular-shaped ring.  The square- and triangular-shaped rings both have a total waveguide length of $L$ and have the same geometric properties given in Table~\ref{Tab:Props}.  In both cases, use of the acoustic vortex wave antenna enables multiple topological modes to be examined using a single aperture design.  Based on the results of Sec.~\ref{Sec:FractModes}, the current analysis will focus on integer modes, since they are the only topological components that contribute to the far-field vortices.  

Figure~\ref{Fig:SquareModes} shows the pressure magnitude for a square-shaped acoustic vortex wave antenna at the first four topological modes.  The magnitude of the radiated pressure field is shown at the source plane ($z/L \! = \! 0$) in (a)--(d), and at $z/L \! = \! 0.5$ in (e)--(h), with the solid line denoting the 3-dB resolution limit given by Eq.~(\ref{Eq:CircRes}).  One of the most noticeable characteristics of the radiated pressure fields illustrated is the distinct difference from the square-shaped ring of the aperture, visible in yellow in (a)--(d), compared with the more complicated vortex patterns that are radiated in (e)--(h).  However, the origins of these complex pressure fields can be seen in the nulls of the source plane created within the boundary formed by the square-shaped ring, and the overall beam pattern extending out beyond the ring.

\begin{figure}[t!]
	\includegraphics[width=0.99\columnwidth, height=0.7\textheight, keepaspectratio]{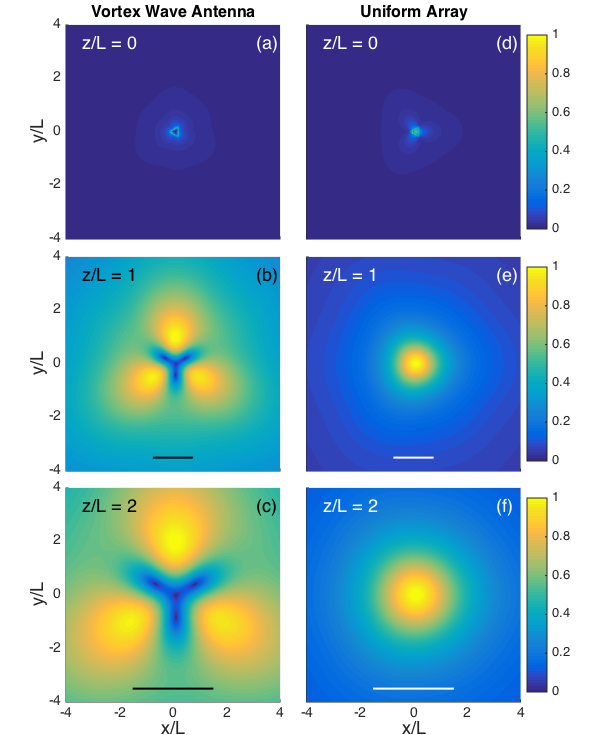}
	\caption{ (Color online) Pressure (magnitude)  obtained analytically using Eq.~(\ref{Eq:Ptot}) for a triangular-shaped arrangement at the mode $m \! = \! 2$ at various distances from the source for (a)--(c) an acoustic vortex wave antenna, and (d)--(f) a uniform acoustic array.  The distances along the $z$ direction from the source plane are normalized by the circumference $L$ of the acoustic antenna.   The corresponding Rayleigh distances from the source for each $z/L$ are tabulated in Table~\ref{Tab:RaylDist}, and the solid lines denote the resolution limit calculated using Eq.~(\ref{Eq:CircRes}). }
	\label{Fig:TriangleMagnitudes}
\end{figure}

In Fig.~\ref{Fig:SquareModes}(e)--(h), the effects of vortex splitting at higher topological modes can also be observed.  Whereas the single vortex at $m \! = \! 1$ remains centered along the beam axis with the peak pressure forming a uniform doughnut-shaped ring (similar to the circular acoustic vortex wave antenna results from Fig.~\ref{Fig:CircMagnitudes}), vortices for progressively higher topological modes form increasingly complex off-axis vortex arrangements.  In particular, it can be observed that at $m \! = \! 3$, the resulting vortex arrangement forms an ``X" shape, which contains sub-resolution detail.  The resolution for the square-shaped ring based the size of these features are tabulated in Table~\ref{Tab:Res}, and give a resolution that is 4-9 times \emph{smaller} than the resolution limit.  A similar effect can be observed at $m \! = \! 4$, and although the center-to-center distances between the vortices extends beyond the resolution limit, the finer detail in the pressure field is smaller than that of the resolution limit denoted by the solid line.

\begin{figure}[t!]
	\includegraphics[width=0.99\columnwidth, height=0.7\textheight, keepaspectratio]{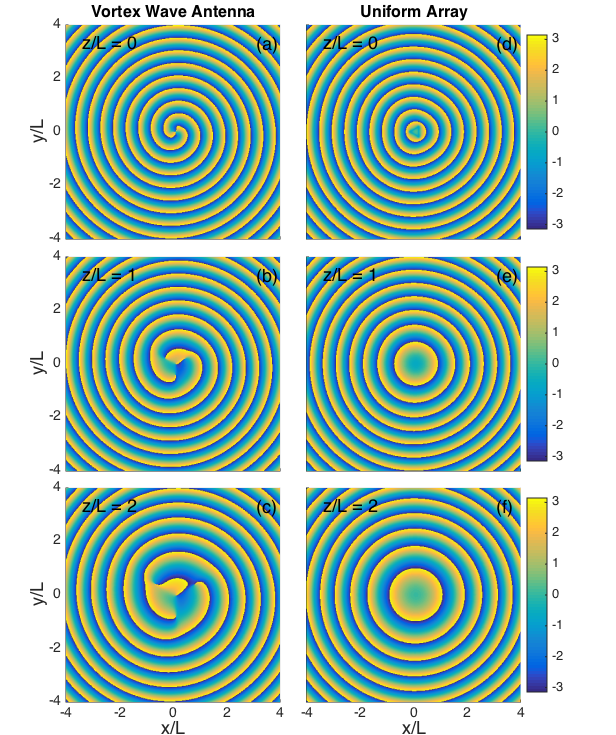}
	\caption{ (Color online) Pressure (phase)  obtained analytically using Eq.~(\ref{Eq:Ptot}) for a triangular-shaped arrangement at the mode $m \! = \! 2$ at various distances from the source for (a)--(c) an acoustic vortex wave antenna, and (d)--(f) a uniform acoustic array.  The distances along the $z$ direction from the source plane are normalized by the circumference $L$ of the acoustic antenna, with the corresponding Rayleigh distances from the source for each $z/L$ are tabulated in Table~\ref{Tab:RaylDist}. }
	\label{Fig:TrianglePhases}
\end{figure}

To further explore the far-field acoustic pressure, the magnitude and phase of the $m \!=\! 3$ mode are presented in Fig.~\ref{Fig:SquareMagnitudes} and \ref{Fig:SquarePhases}, respectively, in panels (a)--(c).  To illustrate the enhanced resolution achieved using the acoustic vortex waves,  results for an equivalent uniform array (using the same square-shaped ring but without the spiral phase created by the acoustic vortex wave antenna) is shown in panels (d)--(f).  The stability of the superresolved ``X" shape can be seen from the results shown in Fig.~\ref{Fig:SquareModes}(g) through well into the far-field in Fig.~\ref{Fig:SquareMagnitudes}(a)--(c), which remains the same except for the overall scaling due to geometric spreading.  By comparison, the uniform array illustrated in Fig.~\ref{Fig:SquareMagnitudes}(d)--(f) achieves a similar resolution as that predicted by Eq.~(\ref{Eq:CircRes}), and the square-shaped features of the ring (present in the source plane) are lost as the beam propagates through the far-field.  Similar results can be observed in the plots of the phase, shown in Fig.~\ref{Fig:SquarePhases}.  In particular, the near-axis phase information retains a detailed, sub-resolution limit structure, whereas the uniform array decays from the square-shaped ring into a circular shape, losing the identifiable detailed information of the particular shape of the source.

Similar results can be also observed for shaped acoustic vortices generated from a triangular-shaped ring.  Figure~\ref{Fig:TriangleModes} shows the pressure magnitude for a triangular-shaped acoustic vortex wave antenna at the first four topological modes, at the source plane ($z/L \! = \! 0$) in panels (a)--(d), and at $z/L \! = \! 0.5$ in panels (e)--(h).  As in the previous case of the square-shaped ring, the solid line denoting the 3-dB resolution limit is given by Eq.~(\ref{Eq:CircRes}), and the radiated pressure fields in panels (e)--(h) create complicated sub-resolution limit vortex patterns which are 5-8 times smaller than the resolution limit, as summarized in Table~\ref{Tab:Res}.  While these radiated pressure fields are distinctly different from the triangular-shaped ring of the aperture, visible in yellow in (a)--(d), similarities can be found in the nulls of the source plane within the boundary formed by the triangle, and the overall beam pattern extending out beyond the ring.

In Fig.~\ref{Fig:TriangleModes}(e)--(h), the effects of vortex splitting at higher topological modes can also be observed.  In a similar manner to the circular and square-shaped acoustic vortex wave antenna results, the single vortex at $m \! = \! 1$ remains centered along the beam axis.  For progressively higher topological modes, increasingly complex off-axis vortex arrangements can be observed in Fig.~\ref{Fig:TriangleModes}(e)--(h).  In particular, it can be observed that at $m \! = \! 2$, the resulting vortex arrangement forms a ``Y" shape, which contains sub-resolution detail.  A similar effect can be observed at $m \! = \! 3$ and $m \! = \! 4$, with finer sub-resolution limit detail in the pressure field occurring within the central triangular region between the vortices.

To further explore the far-field acoustic pressure for the triangular-shaped ring, the magnitude and phase of the $m \!=\! 2$ mode are presented in Fig.~\ref{Fig:TriangleMagnitudes} and \ref{Fig:TrianglePhases}, respectively, in panels (a)--(c).  To illustrate the enhanced resolution achieved using the acoustic vortex waves,  results for an equivalent uniform array are shown in panels (d)--(f).  The uniform array in this case consists of the same triangular-shaped ring, but without the spiral phase created by the acoustic vortex wave antenna.  The stability of the superresolved ``Y" shape can be seen from both the results shown in Fig.~\ref{Fig:TriangleModes}(f) and in Fig.~\ref{Fig:TriangleMagnitudes}(a)--(c), which remains the same except for the overall scaling due to geometric spreading.  By comparison, the uniform array illustrated in Fig.~\ref{Fig:TriangleMagnitudes}(d)--(f) achieves a similar resolution as that predicted by Eq.~(\ref{Eq:CircRes}), and the triangular-shaped features of the ring (present in the source plane) are lost as the beam propagates through the far-field, appearing indistinguishable from a circular aperture.  

Similar results can be observed  in Fig.~\ref{Fig:TrianglePhases} for the phase of the pressure field, for which vortex nodes connected by line discontinuities form a ``Y" pattern near the axis.  In particular, the near-axis phase information retains a detailed yet stable sub-resolution limit ``Y" shape well into the far-field, whereas the uniform array decays from the triangular-shaped ring into a circular shape, losing the identifiable detailed information of the particular shape of the source.

\section{Conclusions} \label{Sec:Conclusions}

In conclusion, an acoustic vortex wave antenna can provide a method for creating shaped acoustic vortices that are stable well into the far-field and enable subdiffraction-limited features without the need for a near-field lens. In this work, a detailed development of the acoustic antenna was presented based on an acoustic leaky-wave antenna.  A key feature of the acoustic antenna is the ability to not only control the wave propagation within the waveguide, but also effectively radiate the acoustic signals to the surrounding medium.   Due to its topological diversity, the acoustic vortex wave antenna is shown to be an ideal means to examine the generation of integer and non-integer topological modes.  Through the use of an acoustic vortex wave antenna and the formulations developed in this work, the decay of fractional topological modes were examined, showing how they change from fractional to integer modes in the far-field.  It was also shown that these integer modes create acoustic vortices that are much smaller than the resolution limit and produce far-field sub-resolution limit features and shapes.

An important aspect of the theoretical formulation developed here is that it is not limited to axisymmetric shapes, and the results of this work were demonstrated for both circular and arbitrarily-shaped acoustic vortex wave antennas.  In both arrangements, it was shown that far-field superresolution could be achieved, and that in particular a non-axisymmetric aperture enabled the creation of complex superresolved off-axis pressure fields due to vortex splitting at higher topological modes, which is useful for communications or particle manipulation applications.  Superresolution using shaped acoustic vortices was demonstrated for the two representative cases of a square- and triangular-shaped ring.  In both cases, superresolved features were demonstrated, with feature sizes 4-9 times smaller than the resolution limit, enabling the creation of shapes including the letter ``X" and ``Y" (for the square- and triangular-shaped rings, respectively), which were stable well into the far-field.  These results were compared with a uniform array, which had the same arrangement of radiating elements but did not generate acoustic vortex waves, and it was shown that these traditional arrays where unable to maintain the relevant detail from the source plane to create sharp features and sub-resolution limit detail.

The results presented here demonstrate how far-field superresolution from an acoustically small source can be achieved through the use of an acoustic vortex wave antenna.  This enables the ability to extend beyond the traditional diffraction-limits on the generation and transmission of acoustic waves, leading to higher precision and control of the far-field pressure.  The improved precision and control of the acoustic pressure using shaped acoustic vortices offers the potential for wide applicability, including acoustic communication and particle manipulation.

\section*{Acknowledgements}
This work was supported by the U.S. Office of Naval Research.


\begin{thebibliography}{71}%
\makeatletter
\providecommand \@ifxundefined [1]{%
 \@ifx{#1\undefined}
}%
\providecommand \@ifnum [1]{%
 \ifnum #1\expandafter \@firstoftwo
 \else \expandafter \@secondoftwo
 \fi
}%
\providecommand \@ifx [1]{%
 \ifx #1\expandafter \@firstoftwo
 \else \expandafter \@secondoftwo
 \fi
}%
\providecommand \natexlab [1]{#1}%
\providecommand \enquote  [1]{``#1''}%
\providecommand \bibnamefont  [1]{#1}%
\providecommand \bibfnamefont [1]{#1}%
\providecommand \citenamefont [1]{#1}%
\providecommand \href@noop [0]{\@secondoftwo}%
\providecommand \href [0]{\begingroup \@sanitize@url \@href}%
\providecommand \@href[1]{\@@startlink{#1}\@@href}%
\providecommand \@@href[1]{\endgroup#1\@@endlink}%
\providecommand \@sanitize@url [0]{\catcode `\\12\catcode `\$12\catcode
  `\&12\catcode `\#12\catcode `\^12\catcode `\_12\catcode `\%12\relax}%
\providecommand \@@startlink[1]{}%
\providecommand \@@endlink[0]{}%
\providecommand \url  [0]{\begingroup\@sanitize@url \@url }%
\providecommand \@url [1]{\endgroup\@href {#1}{\urlprefix }}%
\providecommand \urlprefix  [0]{URL }%
\providecommand \Eprint [0]{\href }%
\providecommand \doibase [0]{http://dx.doi.org/}%
\providecommand \selectlanguage [0]{\@gobble}%
\providecommand \bibinfo  [0]{\@secondoftwo}%
\providecommand \bibfield  [0]{\@secondoftwo}%
\providecommand \translation [1]{[#1]}%
\providecommand \BibitemOpen [0]{}%
\providecommand \bibitemStop [0]{}%
\providecommand \bibitemNoStop [0]{.\EOS\space}%
\providecommand \EOS [0]{\spacefactor3000\relax}%
\providecommand \BibitemShut  [1]{\csname bibitem#1\endcsname}%
\let\auto@bib@innerbib\@empty
\bibitem [{\citenamefont {Cummer}\ \emph {et~al.}(2016)\citenamefont {Cummer},
  \citenamefont {Christensen},\ and\ \citenamefont {Al\`{u}}}]{Cummer2016}%
  \BibitemOpen
  \bibfield  {author} {\bibinfo {author} {\bibfnamefont {S.~A.}\ \bibnamefont
  {Cummer}}, \bibinfo {author} {\bibfnamefont {J.}~\bibnamefont {Christensen}},
  \ and\ \bibinfo {author} {\bibfnamefont {A.}~\bibnamefont {Al\`{u}}},\
  }\bibfield  {title} {\enquote {\bibinfo {title} {Controlling sound with
  acoustic metamaterials},}\ }\href@noop {} {\bibfield  {journal} {\bibinfo
  {journal} {Nat. Rev. Mat.}\ }\textbf {\bibinfo {volume} {1}},\ \bibinfo
  {pages} {16001} (\bibinfo {year} {2016})}\BibitemShut {NoStop}%
\bibitem [{\citenamefont {Ma}\ and\ \citenamefont {Sheng}(2016)}]{Ma2016}%
  \BibitemOpen
  \bibfield  {author} {\bibinfo {author} {\bibfnamefont {G.}~\bibnamefont
  {Ma}}\ and\ \bibinfo {author} {\bibfnamefont {P.}~\bibnamefont {Sheng}},\
  }\bibfield  {title} {\enquote {\bibinfo {title} {Acoustic metamaterials: From
  local resonances to broad horizons},}\ }\href@noop {} {\bibfield  {journal}
  {\bibinfo  {journal} {Sci. Adv.}\ }\textbf {\bibinfo {volume} {2}},\ \bibinfo
  {pages} {e1501595} (\bibinfo {year} {2016})}\BibitemShut {NoStop}%
\bibitem [{\citenamefont {Pendry}(2000)}]{Pendry2000}%
  \BibitemOpen
  \bibfield  {author} {\bibinfo {author} {\bibfnamefont {J.~B.}\ \bibnamefont
  {Pendry}},\ }\bibfield  {title} {\enquote {\bibinfo {title} {Negative
  refraction makes a perfect lens},}\ }\href@noop {} {\bibfield  {journal}
  {\bibinfo  {journal} {Phys. Rev. Lett.}\ }\textbf {\bibinfo {volume} {85}},\
  \bibinfo {pages} {3966--3969} (\bibinfo {year} {2000})}\BibitemShut {NoStop}%
\bibitem [{\citenamefont {Zhang}\ and\ \citenamefont {Liu}(2008)}]{Zhang2008}%
  \BibitemOpen
  \bibfield  {author} {\bibinfo {author} {\bibfnamefont {X.}~\bibnamefont
  {Zhang}}\ and\ \bibinfo {author} {\bibfnamefont {Z.}~\bibnamefont {Liu}},\
  }\bibfield  {title} {\enquote {\bibinfo {title} {Superlenses to overcome the
  diffraction limit},}\ }\href@noop {} {\bibfield  {journal} {\bibinfo
  {journal} {Nat. Mat.}\ }\textbf {\bibinfo {volume} {7}},\ \bibinfo {pages}
  {435--441} (\bibinfo {year} {2008})}\BibitemShut {NoStop}%
\bibitem [{\citenamefont {Zhang}\ \emph {et~al.}(2009)\citenamefont {Zhang},
  \citenamefont {Yin},\ and\ \citenamefont {Fang}}]{Zhang2009}%
  \BibitemOpen
  \bibfield  {author} {\bibinfo {author} {\bibfnamefont {S.}~\bibnamefont
  {Zhang}}, \bibinfo {author} {\bibfnamefont {L.}~\bibnamefont {Yin}}, \ and\
  \bibinfo {author} {\bibfnamefont {N.}~\bibnamefont {Fang}},\ }\bibfield
  {title} {\enquote {\bibinfo {title} {Focusing ultrasound with an acoustic
  metamaterial network},}\ }\href@noop {} {\bibfield  {journal} {\bibinfo
  {journal} {Phys. Rev. Lett.}\ }\textbf {\bibinfo {volume} {102}},\ \bibinfo
  {pages} {194301} (\bibinfo {year} {2009})}\BibitemShut {NoStop}%
\bibitem [{\citenamefont {Zhu}\ \emph {et~al.}(2011)\citenamefont {Zhu},
  \citenamefont {Christensen}, \citenamefont {Jung}, \citenamefont
  {Martin-Moreno}, \citenamefont {Yin}, \citenamefont {Fok},\ and\
  \citenamefont {Zhang}}]{Zhu2011}%
  \BibitemOpen
  \bibfield  {author} {\bibinfo {author} {\bibfnamefont {J.}~\bibnamefont
  {Zhu}}, \bibinfo {author} {\bibfnamefont {J.}~\bibnamefont {Christensen}},
  \bibinfo {author} {\bibfnamefont {J.}~\bibnamefont {Jung}}, \bibinfo {author}
  {\bibfnamefont {L.}~\bibnamefont {Martin-Moreno}}, \bibinfo {author}
  {\bibfnamefont {X.}~\bibnamefont {Yin}}, \bibinfo {author} {\bibfnamefont
  {L.}~\bibnamefont {Fok}}, \ and\ \bibinfo {author} {\bibfnamefont
  {Z.}~\bibnamefont {Zhang}},\ }\bibfield  {title} {\enquote {\bibinfo {title}
  {A holey-structured metamaterial for acoustic deep-subwavelength imaging},}\
  }\href@noop {} {\bibfield  {journal} {\bibinfo  {journal} {Nat. Phys.}\
  }\textbf {\bibinfo {volume} {7}},\ \bibinfo {pages} {52--55} (\bibinfo {year}
  {2011})}\BibitemShut {NoStop}%
\bibitem [{\citenamefont {Garc\'{i}a-Chocano}\ \emph
  {et~al.}(2014)\citenamefont {Garc\'{i}a-Chocano}, \citenamefont
  {Christensen},\ and\ \citenamefont {S\'{a}nchez-Dehesa}}]{GarciaChocano2014}%
  \BibitemOpen
  \bibfield  {author} {\bibinfo {author} {\bibfnamefont {V.~M.}\ \bibnamefont
  {Garc\'{i}a-Chocano}}, \bibinfo {author} {\bibfnamefont {J.}~\bibnamefont
  {Christensen}}, \ and\ \bibinfo {author} {\bibfnamefont {J.}~\bibnamefont
  {S\'{a}nchez-Dehesa}},\ }\bibfield  {title} {\enquote {\bibinfo {title}
  {Negative refraction and energy funneling by hyperbolic materials: an
  experimental demonstration in acoustics},}\ }\href@noop {} {\bibfield
  {journal} {\bibinfo  {journal} {Phys. Rev. Lett.}\ }\textbf {\bibinfo
  {volume} {112}},\ \bibinfo {pages} {144301} (\bibinfo {year}
  {2014})}\BibitemShut {NoStop}%
\bibitem [{\citenamefont {Sheppard}\ and\ \citenamefont
  {Choudhury}(2004)}]{Sheppard2004}%
  \BibitemOpen
  \bibfield  {author} {\bibinfo {author} {\bibfnamefont {C.~J.~R.}\
  \bibnamefont {Sheppard}}\ and\ \bibinfo {author} {\bibfnamefont
  {A.}~\bibnamefont {Choudhury}},\ }\bibfield  {title} {\enquote {\bibinfo
  {title} {Annular pupils, radial polarization and superresolution},}\
  }\href@noop {} {\bibfield  {journal} {\bibinfo  {journal} {Appl. Opt.}\
  }\textbf {\bibinfo {volume} {43}},\ \bibinfo {pages} {4322--4327} (\bibinfo
  {year} {2004})}\BibitemShut {NoStop}%
\bibitem [{\citenamefont {Watanabe}\ \emph {et~al.}(2004)\citenamefont
  {Watanabe}, \citenamefont {Fujii}, \citenamefont {Watanabe}, \citenamefont
  {Toyama},\ and\ \citenamefont {Iketaki}}]{Watanabe2004}%
  \BibitemOpen
  \bibfield  {author} {\bibinfo {author} {\bibfnamefont {T.}~\bibnamefont
  {Watanabe}}, \bibinfo {author} {\bibfnamefont {M.}~\bibnamefont {Fujii}},
  \bibinfo {author} {\bibfnamefont {Y.}~\bibnamefont {Watanabe}}, \bibinfo
  {author} {\bibfnamefont {N.}~\bibnamefont {Toyama}}, \ and\ \bibinfo {author}
  {\bibfnamefont {Y.}~\bibnamefont {Iketaki}},\ }\bibfield  {title} {\enquote
  {\bibinfo {title} {Generation of a doughnut-shaped beam using a spiral phase
  plate},}\ }\href@noop {} {\bibfield  {journal} {\bibinfo  {journal} {Rev.
  Sci. Instrum.}\ }\textbf {\bibinfo {volume} {75}},\ \bibinfo {pages}
  {5131--5135} (\bibinfo {year} {2004})}\BibitemShut {NoStop}%
\bibitem [{\citenamefont {Bokor}\ and\ \citenamefont
  {Davidson}(2007)}]{Bokor2007}%
  \BibitemOpen
  \bibfield  {author} {\bibinfo {author} {\bibfnamefont {N.}~\bibnamefont
  {Bokor}}\ and\ \bibinfo {author} {\bibfnamefont {N.}~\bibnamefont
  {Davidson}},\ }\bibfield  {title} {\enquote {\bibinfo {title} {Tight
  parabolic dark spot with high numerical aperture focusing with a circular
  $\pi$ phase plate},}\ }\href@noop {} {\bibfield  {journal} {\bibinfo
  {journal} {Opt. Commun.}\ }\textbf {\bibinfo {volume} {270}},\ \bibinfo
  {pages} {145--150} (\bibinfo {year} {2007})}\BibitemShut {NoStop}%
\bibitem [{\citenamefont {Nye}\ and\ \citenamefont {Berry}(2000)}]{Nye1974}%
  \BibitemOpen
  \bibfield  {author} {\bibinfo {author} {\bibfnamefont {J.~F.}\ \bibnamefont
  {Nye}}\ and\ \bibinfo {author} {\bibfnamefont {M.~V.}\ \bibnamefont
  {Berry}},\ }\bibfield  {title} {\enquote {\bibinfo {title} {Dislocations in
  wave trains},}\ }\href@noop {} {\bibfield  {journal} {\bibinfo  {journal}
  {Proc. R. Soc. A}\ }\textbf {\bibinfo {volume} {456}},\ \bibinfo {pages}
  {2059--2079} (\bibinfo {year} {2000})}\BibitemShut {NoStop}%
\bibitem [{\citenamefont {Berry}(2004)}]{Berry2004}%
  \BibitemOpen
  \bibfield  {author} {\bibinfo {author} {\bibfnamefont {M.~V.}\ \bibnamefont
  {Berry}},\ }\bibfield  {title} {\enquote {\bibinfo {title} {Optical vortices
  evolving from helicoidal integer and fractional phase steps},}\ }\href@noop
  {} {\bibfield  {journal} {\bibinfo  {journal} {J. Opt. A: Pure Appl. Opt.}\
  }\textbf {\bibinfo {volume} {6}},\ \bibinfo {pages} {259--268} (\bibinfo
  {year} {2004})}\BibitemShut {NoStop}%
\bibitem [{\citenamefont {Allen}\ \emph {et~al.}(1992)\citenamefont {Allen},
  \citenamefont {Beijersbergen}, \citenamefont {Spreeuw},\ and\ \citenamefont
  {Woerdman}}]{Allen1992}%
  \BibitemOpen
  \bibfield  {author} {\bibinfo {author} {\bibfnamefont {L.}~\bibnamefont
  {Allen}}, \bibinfo {author} {\bibfnamefont {M.~W.}\ \bibnamefont
  {Beijersbergen}}, \bibinfo {author} {\bibfnamefont {R.~J.~C.}\ \bibnamefont
  {Spreeuw}}, \ and\ \bibinfo {author} {\bibfnamefont {J.~P.}\ \bibnamefont
  {Woerdman}},\ }\bibfield  {title} {\enquote {\bibinfo {title} {Orbital
  angular momentum of light and the transformation of laguerre-gaussian laser
  modes},}\ }\href@noop {} {\bibfield  {journal} {\bibinfo  {journal} {Phys.
  Rev. A}\ }\textbf {\bibinfo {volume} {45}},\ \bibinfo {pages} {8185--8189}
  (\bibinfo {year} {1992})}\BibitemShut {NoStop}%
\bibitem [{\citenamefont {Yao}\ and\ \citenamefont {Padgett}(2011)}]{Yao2011}%
  \BibitemOpen
  \bibfield  {author} {\bibinfo {author} {\bibfnamefont {A.~M.}\ \bibnamefont
  {Yao}}\ and\ \bibinfo {author} {\bibfnamefont {M.~J.}\ \bibnamefont
  {Padgett}},\ }\bibfield  {title} {\enquote {\bibinfo {title} {Orbital angular
  momentum: origins, behavior and applications},}\ }\href@noop {} {\bibfield
  {journal} {\bibinfo  {journal} {Adv. Opt. Photonics}\ }\textbf {\bibinfo
  {volume} {3}},\ \bibinfo {pages} {161--204} (\bibinfo {year}
  {2011})}\BibitemShut {NoStop}%
\bibitem [{\citenamefont {Wang}\ \emph {et~al.}(2012)\citenamefont {Wang},
  \citenamefont {Yang}, \citenamefont {Fazal}, \citenamefont {Ahmed},
  \citenamefont {Yan}, \citenamefont {Huang}, \citenamefont {Ren},
  \citenamefont {Yue}, \citenamefont {Dolinar}, \citenamefont {Tur},\ and\
  \citenamefont {Willner}}]{Wang2012}%
  \BibitemOpen
  \bibfield  {author} {\bibinfo {author} {\bibfnamefont {J.}~\bibnamefont
  {Wang}}, \bibinfo {author} {\bibfnamefont {J.-Y.}\ \bibnamefont {Yang}},
  \bibinfo {author} {\bibfnamefont {I.~M.}\ \bibnamefont {Fazal}}, \bibinfo
  {author} {\bibfnamefont {N.}~\bibnamefont {Ahmed}}, \bibinfo {author}
  {\bibfnamefont {Y.}~\bibnamefont {Yan}}, \bibinfo {author} {\bibfnamefont
  {H.}~\bibnamefont {Huang}}, \bibinfo {author} {\bibfnamefont
  {Y.}~\bibnamefont {Ren}}, \bibinfo {author} {\bibfnamefont {Y.}~\bibnamefont
  {Yue}}, \bibinfo {author} {\bibfnamefont {S.}~\bibnamefont {Dolinar}},
  \bibinfo {author} {\bibfnamefont {M.}~\bibnamefont {Tur}}, \ and\ \bibinfo
  {author} {\bibfnamefont {A.~E.}\ \bibnamefont {Willner}},\ }\bibfield
  {title} {\enquote {\bibinfo {title} {Terabit free-space data transmission
  employing orbital angular momentum multiplexing},}\ }\href@noop {} {\bibfield
   {journal} {\bibinfo  {journal} {Nature Photon.}\ }\textbf {\bibinfo {volume}
  {6}},\ \bibinfo {pages} {488--496} (\bibinfo {year} {2012})}\BibitemShut
  {NoStop}%
\bibitem [{\citenamefont {Bozinovic}\ \emph {et~al.}(2013)\citenamefont
  {Bozinovic}, \citenamefont {Yue}, \citenamefont {Ren}, \citenamefont {Tur},
  \citenamefont {Kristensen}, \citenamefont {Huang}, \citenamefont {Willner},\
  and\ \citenamefont {Ramachandran}}]{Bozinovic2013}%
  \BibitemOpen
  \bibfield  {author} {\bibinfo {author} {\bibfnamefont {N.}~\bibnamefont
  {Bozinovic}}, \bibinfo {author} {\bibfnamefont {Y.}~\bibnamefont {Yue}},
  \bibinfo {author} {\bibfnamefont {Y.}~\bibnamefont {Ren}}, \bibinfo {author}
  {\bibfnamefont {M.}~\bibnamefont {Tur}}, \bibinfo {author} {\bibfnamefont
  {P.}~\bibnamefont {Kristensen}}, \bibinfo {author} {\bibfnamefont
  {H.}~\bibnamefont {Huang}}, \bibinfo {author} {\bibfnamefont {A.~E.}\
  \bibnamefont {Willner}}, \ and\ \bibinfo {author} {\bibfnamefont
  {S.}~\bibnamefont {Ramachandran}},\ }\bibfield  {title} {\enquote {\bibinfo
  {title} {Terabit-scale orbital angular momentum mode division multiplexing in
  fibers},}\ }\href@noop {} {\bibfield  {journal} {\bibinfo  {journal}
  {Science}\ }\textbf {\bibinfo {volume} {340}},\ \bibinfo {pages} {1545--1548}
  (\bibinfo {year} {2013})}\BibitemShut {NoStop}%
\bibitem [{\citenamefont {Dennis}\ \emph {et~al.}(2009)\citenamefont {Dennis},
  \citenamefont {O'Holleran},\ and\ \citenamefont {Padgett}}]{Dennis2009}%
  \BibitemOpen
  \bibfield  {author} {\bibinfo {author} {\bibfnamefont {M.~R.}\ \bibnamefont
  {Dennis}}, \bibinfo {author} {\bibfnamefont {K.}~\bibnamefont {O'Holleran}},
  \ and\ \bibinfo {author} {\bibfnamefont {M.~J.}\ \bibnamefont {Padgett}},\
  }\bibfield  {title} {\enquote {\bibinfo {title} {Singular optics: optical
  vortices and polarization singularities},}\ }\href@noop {} {\bibfield
  {journal} {\bibinfo  {journal} {Prog. Optics}\ }\textbf {\bibinfo {volume}
  {53}},\ \bibinfo {pages} {293--363} (\bibinfo {year} {2009})}\BibitemShut
  {NoStop}%
\bibitem [{\citenamefont {D'Aguanno}\ \emph {et~al.}(2008)\citenamefont
  {D'Aguanno}, \citenamefont {Mattiucci}, \citenamefont {Bloemer},\ and\
  \citenamefont {Desyatnikov}}]{DAguanno2008}%
  \BibitemOpen
  \bibfield  {author} {\bibinfo {author} {\bibfnamefont {G.}~\bibnamefont
  {D'Aguanno}}, \bibinfo {author} {\bibfnamefont {N.}~\bibnamefont
  {Mattiucci}}, \bibinfo {author} {\bibfnamefont {M.~J.}\ \bibnamefont
  {Bloemer}}, \ and\ \bibinfo {author} {\bibfnamefont {A.}~\bibnamefont
  {Desyatnikov}},\ }\bibfield  {title} {\enquote {\bibinfo {title} {Optical
  vortices during a superresolution process in a metamaterial},}\ }\href@noop
  {} {\bibfield  {journal} {\bibinfo  {journal} {Phys. Rev. A}\ }\textbf
  {\bibinfo {volume} {77}},\ \bibinfo {pages} {043825} (\bibinfo {year}
  {2008})}\BibitemShut {NoStop}%
\bibitem [{\citenamefont {Foo}\ \emph {et~al.}(2005)\citenamefont {Foo},
  \citenamefont {Palacios},\ and\ \citenamefont
  {G.~A.~Swartzlander}}]{Foo2005}%
  \BibitemOpen
  \bibfield  {author} {\bibinfo {author} {\bibfnamefont {G.}~\bibnamefont
  {Foo}}, \bibinfo {author} {\bibfnamefont {D.~M.}\ \bibnamefont {Palacios}}, \
  and\ \bibinfo {author} {\bibfnamefont {Jr.}\ \bibnamefont
  {G.~A.~Swartzlander}},\ }\bibfield  {title} {\enquote {\bibinfo {title}
  {Optical vortex coronagraph},}\ }\href@noop {} {\bibfield  {journal}
  {\bibinfo  {journal} {Opt. Lett.}\ }\textbf {\bibinfo {volume} {30}},\
  \bibinfo {pages} {3308--3310} (\bibinfo {year} {2005})}\BibitemShut {NoStop}%
\bibitem [{\citenamefont {G.~A.~Swartzlander}\ \emph
  {et~al.}(2008)\citenamefont {G.~A.~Swartzlander}, \citenamefont {Ford},
  \citenamefont {Abdul-Malik}, \citenamefont {Close}, \citenamefont {Peters},
  \citenamefont {Palacios},\ and\ \citenamefont {Wilson}}]{Swartzlander2008}%
  \BibitemOpen
  \bibfield  {author} {\bibinfo {author} {\bibfnamefont {Jr.}\ \bibnamefont
  {G.~A.~Swartzlander}}, \bibinfo {author} {\bibfnamefont {E.~L.}\ \bibnamefont
  {Ford}}, \bibinfo {author} {\bibfnamefont {R.~S.}\ \bibnamefont
  {Abdul-Malik}}, \bibinfo {author} {\bibfnamefont {L.~M.}\ \bibnamefont
  {Close}}, \bibinfo {author} {\bibfnamefont {M.~A.}\ \bibnamefont {Peters}},
  \bibinfo {author} {\bibfnamefont {D.~M.}\ \bibnamefont {Palacios}}, \ and\
  \bibinfo {author} {\bibfnamefont {D.~W.}\ \bibnamefont {Wilson}},\ }\bibfield
   {title} {\enquote {\bibinfo {title} {Astronomical demonstration of an
  optical vortex coronagraph},}\ }\href@noop {} {\bibfield  {journal} {\bibinfo
   {journal} {Opt. Express}\ }\textbf {\bibinfo {volume} {16}},\ \bibinfo
  {pages} {10200--10207} (\bibinfo {year} {2008})}\BibitemShut {NoStop}%
\bibitem [{\citenamefont {Brasselet}\ \emph {et~al.}(2013)\citenamefont
  {Brasselet}, \citenamefont {Gervinskas}, \citenamefont {Seniutinas},\ and\
  \citenamefont {Juodkazis}}]{Brasselet2013}%
  \BibitemOpen
  \bibfield  {author} {\bibinfo {author} {\bibfnamefont {E.}~\bibnamefont
  {Brasselet}}, \bibinfo {author} {\bibfnamefont {G.}~\bibnamefont
  {Gervinskas}}, \bibinfo {author} {\bibfnamefont {G.}~\bibnamefont
  {Seniutinas}}, \ and\ \bibinfo {author} {\bibfnamefont {S.}~\bibnamefont
  {Juodkazis}},\ }\bibfield  {title} {\enquote {\bibinfo {title} {Topological
  shaping of light by closed-path nanoslits},}\ }\href@noop {} {\bibfield
  {journal} {\bibinfo  {journal} {Phys. Rev. Lett.}\ }\textbf {\bibinfo
  {volume} {111}},\ \bibinfo {pages} {193901} (\bibinfo {year}
  {2013})}\BibitemShut {NoStop}%
\bibitem [{\citenamefont {Hickman}\ \emph {et~al.}(2010)\citenamefont
  {Hickman}, \citenamefont {Fonseca}, \citenamefont {Soares},\ and\
  \citenamefont {Ch\'{a}vez-Cerda}}]{Hickman2010}%
  \BibitemOpen
  \bibfield  {author} {\bibinfo {author} {\bibfnamefont {J.~M.}\ \bibnamefont
  {Hickmann}}, \bibinfo {author} {\bibfnamefont {E.~J.~S.}\ \bibnamefont
  {Fonseca}}, \bibinfo {author} {\bibfnamefont {W.~C.}\ \bibnamefont {Soares}},
  \ and\ \bibinfo {author} {\bibfnamefont {S.}~\bibnamefont
  {Ch\'{a}vez-Cerda}},\ }\bibfield  {title} {\enquote {\bibinfo {title}
  {Unveiling a truncated optical lattice associated with a triangular aperture
  using light's orbital angular momentum},}\ }\href@noop {} {\bibfield
  {journal} {\bibinfo  {journal} {Phys. Rev. Lett.}\ }\textbf {\bibinfo
  {volume} {105}},\ \bibinfo {pages} {053904} (\bibinfo {year}
  {2010})}\BibitemShut {NoStop}%
\bibitem [{\citenamefont {Amaral}\ \emph {et~al.}(2013)\citenamefont {Amaral},
  \citenamefont {Falc{\~a}o-Filho},\ and\ \citenamefont
  {de~Ara\'{u}jo}}]{Amaral2013}%
  \BibitemOpen
  \bibfield  {author} {\bibinfo {author} {\bibfnamefont {A.~M.}\ \bibnamefont
  {Amaral}}, \bibinfo {author} {\bibfnamefont {E.~L.}\ \bibnamefont
  {Falc{\~a}o-Filho}}, \ and\ \bibinfo {author} {\bibfnamefont {C.~B.}\
  \bibnamefont {de~Ara\'{u}jo}},\ }\bibfield  {title} {\enquote {\bibinfo
  {title} {Shaping optical beams with topological charge},}\ }\href@noop {}
  {\bibfield  {journal} {\bibinfo  {journal} {Opt. Lett.}\ }\textbf {\bibinfo
  {volume} {38}},\ \bibinfo {pages} {1579--1581} (\bibinfo {year}
  {2013})}\BibitemShut {NoStop}%
\bibitem [{\citenamefont {Amaral}\ \emph {et~al.}(2014)\citenamefont {Amaral},
  \citenamefont {Falc{\~a}o-Filho},\ and\ \citenamefont
  {de~Ara\'{u}jo}}]{Amaral2014}%
  \BibitemOpen
  \bibfield  {author} {\bibinfo {author} {\bibfnamefont {A.~M.}\ \bibnamefont
  {Amaral}}, \bibinfo {author} {\bibfnamefont {E.~L.}\ \bibnamefont
  {Falc{\~a}o-Filho}}, \ and\ \bibinfo {author} {\bibfnamefont {C.~B.}\
  \bibnamefont {de~Ara\'{u}jo}},\ }\bibfield  {title} {\enquote {\bibinfo
  {title} {Characterization of topological charge and orbital angular momentum
  of shaped optical vortices},}\ }\href@noop {} {\bibfield  {journal} {\bibinfo
   {journal} {Opt. Express}\ }\textbf {\bibinfo {volume} {22}},\ \bibinfo
  {pages} {30315--30324} (\bibinfo {year} {2014})}\BibitemShut {NoStop}%
\bibitem [{\citenamefont {Thomas}\ and\ \citenamefont
  {Marchiano}(2003)}]{Thomas2003}%
  \BibitemOpen
  \bibfield  {author} {\bibinfo {author} {\bibfnamefont {J.-L.}\ \bibnamefont
  {Thomas}}\ and\ \bibinfo {author} {\bibfnamefont {R.}~\bibnamefont
  {Marchiano}},\ }\bibfield  {title} {\enquote {\bibinfo {title} {Pseudo
  angular momentum and topological charge conservation for nonlinear acoustical
  vortices},}\ }\href@noop {} {\bibfield  {journal} {\bibinfo  {journal} {Phys.
  Rev. Lett.}\ }\textbf {\bibinfo {volume} {91}},\ \bibinfo {pages} {244302}
  (\bibinfo {year} {2003})}\BibitemShut {NoStop}%
\bibitem [{\citenamefont {Wilson}\ and\ \citenamefont
  {Padgett}(2010)}]{Wilson2010}%
  \BibitemOpen
  \bibfield  {author} {\bibinfo {author} {\bibfnamefont {C.}~\bibnamefont
  {Wilson}}\ and\ \bibinfo {author} {\bibfnamefont {M.~J.}\ \bibnamefont
  {Padgett}},\ }\bibfield  {title} {\enquote {\bibinfo {title} {A polyphonic
  acoustic vortex and its complementary chords},}\ }\href@noop {} {\bibfield
  {journal} {\bibinfo  {journal} {New. J. Phys.}\ }\textbf {\bibinfo {volume}
  {12}},\ \bibinfo {pages} {023018} (\bibinfo {year} {2010})}\BibitemShut
  {NoStop}%
\bibitem [{\citenamefont {Hefner}\ and\ \citenamefont
  {Marston}(1998)}]{Hefner1998}%
  \BibitemOpen
  \bibfield  {author} {\bibinfo {author} {\bibfnamefont {B.~T.}\ \bibnamefont
  {Hefner}}\ and\ \bibinfo {author} {\bibfnamefont {P.~L.}\ \bibnamefont
  {Marston}},\ }\bibfield  {title} {\enquote {\bibinfo {title} {Acoustical
  helicoidal waves and {L}aguerre-{G}aussian beams: applications to scattering
  and to angular momentum transport},}\ }in\ \href@noop {} {\emph {\bibinfo
  {booktitle} {Proceedings of the 16th International Congress on Acoustics}}},\
  \bibinfo {editor} {edited by\ \bibinfo {editor} {\bibfnamefont {P.~K.}\
  \bibnamefont {Kuhl}}\ and\ \bibinfo {editor} {\bibfnamefont {L.~A.}\
  \bibnamefont {Crum}}}\ (\bibinfo  {publisher} {ASA},\ \bibinfo {address}
  {Seattle},\ \bibinfo {year} {1998})\ pp.\ \bibinfo {pages}
  {1921--1922}\BibitemShut {NoStop}%
\bibitem [{\citenamefont {Hong}\ \emph {et~al.}(2015)\citenamefont {Hong},
  \citenamefont {Zhang},\ and\ \citenamefont {Drinkwater}}]{Hong2015}%
  \BibitemOpen
  \bibfield  {author} {\bibinfo {author} {\bibfnamefont {Z.~Y.}\ \bibnamefont
  {Hong}}, \bibinfo {author} {\bibfnamefont {J.}~\bibnamefont {Zhang}}, \ and\
  \bibinfo {author} {\bibfnamefont {B.~W.}\ \bibnamefont {Drinkwater}},\
  }\bibfield  {title} {\enquote {\bibinfo {title} {On the radiation force
  fields of fractional-order acoustic vortices},}\ }\href@noop {} {\bibfield
  {journal} {\bibinfo  {journal} {EPL}\ }\textbf {\bibinfo {volume} {110}},\
  \bibinfo {pages} {14002} (\bibinfo {year} {2015})}\BibitemShut {NoStop}%
\bibitem [{\citenamefont {Hefner}\ and\ \citenamefont
  {Marston}(1999)}]{Hefner1999}%
  \BibitemOpen
  \bibfield  {author} {\bibinfo {author} {\bibfnamefont {B.~T.}\ \bibnamefont
  {Hefner}}\ and\ \bibinfo {author} {\bibfnamefont {P.~L.}\ \bibnamefont
  {Marston}},\ }\bibfield  {title} {\enquote {\bibinfo {title} {An acoustical
  helicoidal wave transducer with applications for the alignment of ultrasonic
  and underwater systems},}\ }\href@noop {} {\bibfield  {journal} {\bibinfo
  {journal} {J. Acoust. Soc. Am.}\ }\textbf {\bibinfo {volume} {106}},\
  \bibinfo {pages} {3313--3316} (\bibinfo {year} {1999})}\BibitemShut {NoStop}%
\bibitem [{\citenamefont {Marzo}\ \emph {et~al.}(2015)\citenamefont {Marzo},
  \citenamefont {Seah}, \citenamefont {Drinkwater}, \citenamefont {Sahoo},
  \citenamefont {Long},\ and\ \citenamefont {Subramanian}}]{Marzo2015}%
  \BibitemOpen
  \bibfield  {author} {\bibinfo {author} {\bibfnamefont {A.}~\bibnamefont
  {Marzo}}, \bibinfo {author} {\bibfnamefont {S.~A.}\ \bibnamefont {Seah}},
  \bibinfo {author} {\bibfnamefont {B.~W.}\ \bibnamefont {Drinkwater}},
  \bibinfo {author} {\bibfnamefont {D.~R.}\ \bibnamefont {Sahoo}}, \bibinfo
  {author} {\bibfnamefont {B.}~\bibnamefont {Long}}, \ and\ \bibinfo {author}
  {\bibfnamefont {S.}~\bibnamefont {Subramanian}},\ }\bibfield  {title}
  {\enquote {\bibinfo {title} {Holographic acoustic elements for manipulation
  of levitated objects},}\ }\href@noop {} {\bibfield  {journal} {\bibinfo
  {journal} {Nat. Commun.}\ }\textbf {\bibinfo {volume} {6}},\ \bibinfo {pages}
  {8661} (\bibinfo {year} {2015})}\BibitemShut {NoStop}%
\bibitem [{\citenamefont {Skeldon}\ \emph {et~al.}(2008)\citenamefont
  {Skeldon}, \citenamefont {Wilson}, \citenamefont {Edgar},\ and\ \citenamefont
  {Padgett}}]{Skeldon2008}%
  \BibitemOpen
  \bibfield  {author} {\bibinfo {author} {\bibfnamefont {K.~D.}\ \bibnamefont
  {Skeldon}}, \bibinfo {author} {\bibfnamefont {C.}~\bibnamefont {Wilson}},
  \bibinfo {author} {\bibfnamefont {M.}~\bibnamefont {Edgar}}, \ and\ \bibinfo
  {author} {\bibfnamefont {M.~J.}\ \bibnamefont {Padgett}},\ }\bibfield
  {title} {\enquote {\bibinfo {title} {An acoustic spanner and its associated
  rotational {D}oppler shift},}\ }\href@noop {} {\bibfield  {journal} {\bibinfo
   {journal} {New. J. Phys.}\ }\textbf {\bibinfo {volume} {10}},\ \bibinfo
  {pages} {013018} (\bibinfo {year} {2008})}\BibitemShut {NoStop}%
\bibitem [{\citenamefont {Volke-Sep\'{u}lveda}\ \emph
  {et~al.}(2008)\citenamefont {Volke-Sep\'{u}lveda}, \citenamefont
  {Santill\'{a}n},\ and\ \citenamefont {Boullosa}}]{VolkeSepulveda2008}%
  \BibitemOpen
  \bibfield  {author} {\bibinfo {author} {\bibfnamefont {K.}~\bibnamefont
  {Volke-Sep\'{u}lveda}}, \bibinfo {author} {\bibfnamefont {A.~O.}\
  \bibnamefont {Santill\'{a}n}}, \ and\ \bibinfo {author} {\bibfnamefont
  {R.~R.}\ \bibnamefont {Boullosa}},\ }\bibfield  {title} {\enquote {\bibinfo
  {title} {Transfer of angular momentum to matter from acoustical vortices in
  free space},}\ }\href@noop {} {\bibfield  {journal} {\bibinfo  {journal}
  {Phys. Rev. Lett.}\ }\textbf {\bibinfo {volume} {100}},\ \bibinfo {pages}
  {024302} (\bibinfo {year} {2008})}\BibitemShut {NoStop}%
\bibitem [{\citenamefont {Thomas}\ \emph {et~al.}(2010)\citenamefont {Thomas},
  \citenamefont {Brunet},\ and\ \citenamefont {Coulouvrat}}]{Thomas2010}%
  \BibitemOpen
  \bibfield  {author} {\bibinfo {author} {\bibfnamefont {J.-L.}\ \bibnamefont
  {Thomas}}, \bibinfo {author} {\bibfnamefont {T.}~\bibnamefont {Brunet}}, \
  and\ \bibinfo {author} {\bibfnamefont {F.}~\bibnamefont {Coulouvrat}},\
  }\bibfield  {title} {\enquote {\bibinfo {title} {Generalization of helicoidal
  beams for short pulses},}\ }\href@noop {} {\bibfield  {journal} {\bibinfo
  {journal} {Phys. Rev. E}\ }\textbf {\bibinfo {volume} {81}},\ \bibinfo
  {pages} {016601} (\bibinfo {year} {2010})}\BibitemShut {NoStop}%
\bibitem [{\citenamefont {Marston}(2008)}]{Marston2008}%
  \BibitemOpen
  \bibfield  {author} {\bibinfo {author} {\bibfnamefont {P.~L.}\ \bibnamefont
  {Marston}},\ }\bibfield  {title} {\enquote {\bibinfo {title} {Scattering of a
  {B}essel beam by a sphere: {II}. {H}elicoidal case and spherical shell
  example},}\ }\href@noop {} {\bibfield  {journal} {\bibinfo  {journal} {J.
  Acoust. Soc. Am.}\ }\textbf {\bibinfo {volume} {124}},\ \bibinfo {pages}
  {2905--2910} (\bibinfo {year} {2008})}\BibitemShut {NoStop}%
\bibitem [{\citenamefont {Mitri}(2011)}]{Mitri2011}%
  \BibitemOpen
  \bibfield  {author} {\bibinfo {author} {\bibfnamefont {F.~G.}\ \bibnamefont
  {Mitri}},\ }\bibfield  {title} {\enquote {\bibinfo {title} {Acoustic beam
  interaction with a rigid sphere: The case of a first-order non-diffracting
  bessel trigonometric beam},}\ }\href@noop {} {\bibfield  {journal} {\bibinfo
  {journal} {J. Sound Vib.}\ }\textbf {\bibinfo {volume} {330}},\ \bibinfo
  {pages} {6053--6060} (\bibinfo {year} {2011})}\BibitemShut {NoStop}%
\bibitem [{\citenamefont {Marston}(2009{\natexlab{a}})}]{Marston2009}%
  \BibitemOpen
  \bibfield  {author} {\bibinfo {author} {\bibfnamefont {P.~L.}\ \bibnamefont
  {Marston}},\ }\bibfield  {title} {\enquote {\bibinfo {title} {Radiation force
  of a helicoidal {B}essel beam on a sphere},}\ }\href@noop {} {\bibfield
  {journal} {\bibinfo  {journal} {J. Acoust. Soc. Am.}\ }\textbf {\bibinfo
  {volume} {125}},\ \bibinfo {pages} {3529--3547} (\bibinfo {year}
  {2009}{\natexlab{a}})}\BibitemShut {NoStop}%
\bibitem [{\citenamefont {Berry}(2005)}]{Berry2005}%
  \BibitemOpen
  \bibfield  {author} {\bibinfo {author} {\bibfnamefont {M.~V.}\ \bibnamefont
  {Berry}},\ }\bibfield  {title} {\enquote {\bibinfo {title} {Phase vortex
  spirals},}\ }\href@noop {} {\bibfield  {journal} {\bibinfo  {journal} {J.
  Phys. A: Math Gen.}\ }\textbf {\bibinfo {volume} {38}},\ \bibinfo {pages}
  {L745--L751} (\bibinfo {year} {2005})}\BibitemShut {NoStop}%
\bibitem [{\citenamefont {Kotlyar}\ \emph {et~al.}(2006)\citenamefont
  {Kotlyar}, \citenamefont {Khonina}, \citenamefont {Kovalev},\ and\
  \citenamefont {Soifer}}]{Kotlyar2006}%
  \BibitemOpen
  \bibfield  {author} {\bibinfo {author} {\bibfnamefont {V.~V.}\ \bibnamefont
  {Kotlyar}}, \bibinfo {author} {\bibfnamefont {S.~N.}\ \bibnamefont
  {Khonina}}, \bibinfo {author} {\bibfnamefont {A.~A.}\ \bibnamefont
  {Kovalev}}, \ and\ \bibinfo {author} {\bibfnamefont {V.~A.}\ \bibnamefont
  {Soifer}},\ }\bibfield  {title} {\enquote {\bibinfo {title} {Diffraction of a
  plane, finite-radius wave by a spiral phase plate},}\ }\href@noop {}
  {\bibfield  {journal} {\bibinfo  {journal} {Opt. Lett.}\ }\textbf {\bibinfo
  {volume} {31}},\ \bibinfo {pages} {1597--1599} (\bibinfo {year}
  {2006})}\BibitemShut {NoStop}%
\bibitem [{\citenamefont {Vasnetsov}\ \emph {et~al.}(1998)\citenamefont
  {Vasnetsov}, \citenamefont {Basistiy},\ and\ \citenamefont
  {Soskin}}]{Vasnetsov1998}%
  \BibitemOpen
  \bibfield  {author} {\bibinfo {author} {\bibfnamefont {M.~V.}\ \bibnamefont
  {Vasnetsov}}, \bibinfo {author} {\bibfnamefont {I.~V.}\ \bibnamefont
  {Basistiy}}, \ and\ \bibinfo {author} {\bibfnamefont {M.~S.}\ \bibnamefont
  {Soskin}},\ }\bibfield  {title} {\enquote {\bibinfo {title} {Free-space
  evolution of monochromatic mixed screw-edge wavefront dislocations},}\
  }\href@noop {} {\bibfield  {journal} {\bibinfo  {journal} {Proc. SPIE}\
  }\textbf {\bibinfo {volume} {3487}},\ \bibinfo {pages} {29--33} (\bibinfo
  {year} {1998})}\BibitemShut {NoStop}%
\bibitem [{\citenamefont {Lee}\ \emph {et~al.}(2004)\citenamefont {Lee},
  \citenamefont {Yuan},\ and\ \citenamefont {Dholakia}}]{Lee2004}%
  \BibitemOpen
  \bibfield  {author} {\bibinfo {author} {\bibfnamefont {W.~M.}\ \bibnamefont
  {Lee}}, \bibinfo {author} {\bibfnamefont {X.-C.}\ \bibnamefont {Yuan}}, \
  and\ \bibinfo {author} {\bibfnamefont {K.}~\bibnamefont {Dholakia}},\
  }\bibfield  {title} {\enquote {\bibinfo {title} {Experimental observation of
  optical vortex evolution in a {G}aussian beam with an embedded fractional
  phase step},}\ }\href@noop {} {\bibfield  {journal} {\bibinfo  {journal}
  {Opt. Commun.}\ }\textbf {\bibinfo {volume} {239}},\ \bibinfo {pages}
  {129--135} (\bibinfo {year} {2004})}\BibitemShut {NoStop}%
\bibitem [{\citenamefont {Tao}\ \emph {et~al.}(2005)\citenamefont {Tao},
  \citenamefont {Yuan},\ and\ \citenamefont {Lin}}]{Tao2005}%
  \BibitemOpen
  \bibfield  {author} {\bibinfo {author} {\bibfnamefont {S.~H.}\ \bibnamefont
  {Tao}}, \bibinfo {author} {\bibfnamefont {X.-C.}\ \bibnamefont {Yuan}}, \
  and\ \bibinfo {author} {\bibfnamefont {J.}~\bibnamefont {Lin}},\ }\bibfield
  {title} {\enquote {\bibinfo {title} {Fractional optical vortex beam induced
  rotation of particles},}\ }\href@noop {} {\bibfield  {journal} {\bibinfo
  {journal} {Opt. Express}\ }\textbf {\bibinfo {volume} {13}},\ \bibinfo
  {pages} {7726--7731} (\bibinfo {year} {2005})}\BibitemShut {NoStop}%
\bibitem [{\citenamefont {Garcia-Gracia}\ and\ \citenamefont
  {Guti\'{e}rrez-Vega}(2009)}]{GarciaGracia2009}%
  \BibitemOpen
  \bibfield  {author} {\bibinfo {author} {\bibfnamefont {H.}~\bibnamefont
  {Garcia-Gracia}}\ and\ \bibinfo {author} {\bibfnamefont {J.~C.}\ \bibnamefont
  {Guti\'{e}rrez-Vega}},\ }\bibfield  {title} {\enquote {\bibinfo {title}
  {Diffraction of plane waves by finite-radius sprial phase plates of integer
  and fractional topological charge},}\ }\href@noop {} {\bibfield  {journal}
  {\bibinfo  {journal} {J. Opt. Soc. Am. A}\ }\textbf {\bibinfo {volume}
  {26}},\ \bibinfo {pages} {794--803} (\bibinfo {year} {2009})}\BibitemShut
  {NoStop}%
\bibitem [{\citenamefont {Marston}(2009{\natexlab{b}})}]{Marston2009a}%
  \BibitemOpen
  \bibfield  {author} {\bibinfo {author} {\bibfnamefont {P.~L.}\ \bibnamefont
  {Marston}},\ }\bibfield  {title} {\enquote {\bibinfo {title}
  {Self-reconstruction property of fractional bessel beams: comment},}\
  }\href@noop {} {\bibfield  {journal} {\bibinfo  {journal} {J. Opt. Soc. Am.
  A}\ }\textbf {\bibinfo {volume} {26}},\ \bibinfo {pages} {2181} (\bibinfo
  {year} {2009}{\natexlab{b}})}\BibitemShut {NoStop}%
\bibitem [{\citenamefont {Vyas}\ \emph {et~al.}(2010)\citenamefont {Vyas},
  \citenamefont {Kumar},\ and\ \citenamefont {Senthilkumaran}}]{Vyas2010}%
  \BibitemOpen
  \bibfield  {author} {\bibinfo {author} {\bibfnamefont {S.}~\bibnamefont
  {Vyas}}, \bibinfo {author} {\bibfnamefont {R.}~\bibnamefont {Kumar}}, \ and\
  \bibinfo {author} {\bibfnamefont {S.~P.}\ \bibnamefont {Senthilkumaran}},\
  }\bibfield  {title} {\enquote {\bibinfo {title} {Fractional vortex lens},}\
  }\href@noop {} {\bibfield  {journal} {\bibinfo  {journal} {Opt. Laser
  Technol.}\ }\textbf {\bibinfo {volume} {42}},\ \bibinfo {pages} {878--882}
  (\bibinfo {year} {2010})}\BibitemShut {NoStop}%
\bibitem [{\citenamefont {Al-Bassam}\ \emph {et~al.}(2014)\citenamefont
  {Al-Bassam}, \citenamefont {Salem},\ and\ \citenamefont
  {Caloz}}]{AlBassam2014}%
  \BibitemOpen
  \bibfield  {author} {\bibinfo {author} {\bibfnamefont {A.}~\bibnamefont
  {Al-Bassam}}, \bibinfo {author} {\bibfnamefont {M.~A.}\ \bibnamefont
  {Salem}}, \ and\ \bibinfo {author} {\bibfnamefont {C.}~\bibnamefont
  {Caloz}},\ }\bibfield  {title} {\enquote {\bibinfo {title} {Vortex beam
  generation using circular leaky-wave antenna},}\ }\bibinfo {organization}
  {Antennas and Propagation Society International Symposium (APSURSI)}\
  (\bibinfo  {publisher} {IEEE},\ \bibinfo {address} {Memphis, TN},\ \bibinfo
  {year} {2014})\ pp.\ \bibinfo {pages} {1792--1793}\BibitemShut {NoStop}%
\bibitem [{\citenamefont {Jackson}\ \emph {et~al.}(2012)\citenamefont
  {Jackson}, \citenamefont {Caloz},\ and\ \citenamefont {Itoh}}]{Jackson2012}%
  \BibitemOpen
  \bibfield  {author} {\bibinfo {author} {\bibfnamefont {D.~R.}\ \bibnamefont
  {Jackson}}, \bibinfo {author} {\bibfnamefont {C.}~\bibnamefont {Caloz}}, \
  and\ \bibinfo {author} {\bibfnamefont {T.}~\bibnamefont {Itoh}},\ }\bibfield
  {title} {\enquote {\bibinfo {title} {Leaky-wave antennas},}\ }\href@noop {}
  {\bibfield  {journal} {\bibinfo  {journal} {Proc. IEEE}\ }\textbf {\bibinfo
  {volume} {100}},\ \bibinfo {pages} {2194--2206} (\bibinfo {year}
  {2012})}\BibitemShut {NoStop}%
\bibitem [{\citenamefont {Monticone}\ and\ \citenamefont
  {Al\`{u}}(2015)}]{Monticone2015}%
  \BibitemOpen
  \bibfield  {author} {\bibinfo {author} {\bibfnamefont {F.}~\bibnamefont
  {Monticone}}\ and\ \bibinfo {author} {\bibfnamefont {A.}~\bibnamefont
  {Al\`{u}}},\ }\bibfield  {title} {\enquote {\bibinfo {title} {Leaky-wave
  theory, techniques and applications: from microwaves to visible
  frequencies},}\ }\href@noop {} {\bibfield  {journal} {\bibinfo  {journal}
  {Proc. IEEE}\ }\textbf {\bibinfo {volume} {103}},\ \bibinfo {pages}
  {793--821} (\bibinfo {year} {2015})}\BibitemShut {NoStop}%
\bibitem [{\citenamefont {Grbic}\ and\ \citenamefont
  {Eleftheriades}(2002)}]{Grbic2002}%
  \BibitemOpen
  \bibfield  {author} {\bibinfo {author} {\bibfnamefont {A.}~\bibnamefont
  {Grbic}}\ and\ \bibinfo {author} {\bibfnamefont {G.~V.}\ \bibnamefont
  {Eleftheriades}},\ }\bibfield  {title} {\enquote {\bibinfo {title}
  {Experimental verification of backward-wave radiation from a negative
  refractive index metamaterial},}\ }\href@noop {} {\bibfield  {journal}
  {\bibinfo  {journal} {J. Appl. Phys.}\ }\textbf {\bibinfo {volume} {92}},\
  \bibinfo {pages} {5930--5935} (\bibinfo {year} {2002})}\BibitemShut {NoStop}%
\bibitem [{\citenamefont {Liu}\ \emph {et~al.}(2002)\citenamefont {Liu},
  \citenamefont {Caloz},\ and\ \citenamefont {Itoh}}]{Liu2002}%
  \BibitemOpen
  \bibfield  {author} {\bibinfo {author} {\bibfnamefont {L.}~\bibnamefont
  {Liu}}, \bibinfo {author} {\bibfnamefont {C.}~\bibnamefont {Caloz}}, \ and\
  \bibinfo {author} {\bibfnamefont {T.}~\bibnamefont {Itoh}},\ }\bibfield
  {title} {\enquote {\bibinfo {title} {Dominant mode leaky-wave antenna with
  backfire-to-endfire scanning capability},}\ }\href@noop {} {\bibfield
  {journal} {\bibinfo  {journal} {Electron. Lett.}\ }\textbf {\bibinfo {volume}
  {38}},\ \bibinfo {pages} {1414---1416} (\bibinfo {year} {2002})}\BibitemShut
  {NoStop}%
\bibitem [{\citenamefont {Caloz}\ and\ \citenamefont {Itoh}(2004)}]{Caloz2004}%
  \BibitemOpen
  \bibfield  {author} {\bibinfo {author} {\bibfnamefont {C.}~\bibnamefont
  {Caloz}}\ and\ \bibinfo {author} {\bibfnamefont {T.}~\bibnamefont {Itoh}},\
  }\bibfield  {title} {\enquote {\bibinfo {title} {Array factor approach of
  leaky-wave antennas and application to 1-{D}/2-{D} composite
  right/left-handed ({CRLH}) structures},}\ }\href@noop {} {\bibfield
  {journal} {\bibinfo  {journal} {IEEE Microw. Wirel. Compon. Lett.}\ }\textbf
  {\bibinfo {volume} {14}},\ \bibinfo {pages} {274--276} (\bibinfo {year}
  {2004})}\BibitemShut {NoStop}%
\bibitem [{\citenamefont {Lim}\ \emph {et~al.}(2004)\citenamefont {Lim},
  \citenamefont {Caloz},\ and\ \citenamefont {Itoh}}]{Lim2004}%
  \BibitemOpen
  \bibfield  {author} {\bibinfo {author} {\bibfnamefont {S.}~\bibnamefont
  {Lim}}, \bibinfo {author} {\bibfnamefont {C.}~\bibnamefont {Caloz}}, \ and\
  \bibinfo {author} {\bibfnamefont {T.}~\bibnamefont {Itoh}},\ }\bibfield
  {title} {\enquote {\bibinfo {title} {Metamaterial-based electronically
  controlled tranmission-line structure as a novel leaky-wave antenna with
  tunable radiation angle and beamwidth},}\ }\href@noop {} {\bibfield
  {journal} {\bibinfo  {journal} {IEEE Trans. Microw. Theory Tech.}\ }\textbf
  {\bibinfo {volume} {52}},\ \bibinfo {pages} {2678--2690} (\bibinfo {year}
  {2004})}\BibitemShut {NoStop}%
\bibitem [{\citenamefont {Lai}\ \emph {et~al.}(2004)\citenamefont {Lai},
  \citenamefont {Caloz},\ and\ \citenamefont {Itoh}}]{Lai2004}%
  \BibitemOpen
  \bibfield  {author} {\bibinfo {author} {\bibfnamefont {A.}~\bibnamefont
  {Lai}}, \bibinfo {author} {\bibfnamefont {C.}~\bibnamefont {Caloz}}, \ and\
  \bibinfo {author} {\bibfnamefont {T.}~\bibnamefont {Itoh}},\ }\bibfield
  {title} {\enquote {\bibinfo {title} {Composite right/left-handed transmission
  line metamaterials},}\ }\href@noop {} {\bibfield  {journal} {\bibinfo
  {journal} {IEEE Microwave Mag.}\ }\textbf {\bibinfo {volume} {5}},\ \bibinfo
  {pages} {34--50} (\bibinfo {year} {2004})}\BibitemShut {NoStop}%
\bibitem [{\citenamefont {Lai}\ \emph {et~al.}(2006)\citenamefont {Lai},
  \citenamefont {Leong},\ and\ \citenamefont {Itoh}}]{Lai2006}%
  \BibitemOpen
  \bibfield  {author} {\bibinfo {author} {\bibfnamefont {A.}~\bibnamefont
  {Lai}}, \bibinfo {author} {\bibfnamefont {K.}~\bibnamefont {Leong}}, \ and\
  \bibinfo {author} {\bibfnamefont {T.}~\bibnamefont {Itoh}},\ }\bibfield
  {title} {\enquote {\bibinfo {title} {Leaky-wave steering in a two-dimensional
  metamaterial structure using wave interaction excitation},}\ }\href@noop {}
  {\bibfield  {journal} {\bibinfo  {journal} {IEEE}\ ,\ \bibinfo {pages}
  {1643--1646}} (\bibinfo {year} {2006})}\BibitemShut {NoStop}%
\bibitem [{\citenamefont {Caloz}\ \emph {et~al.}(2008)\citenamefont {Caloz},
  \citenamefont {Itoh},\ and\ \citenamefont {Rennings}}]{Caloz2008}%
  \BibitemOpen
  \bibfield  {author} {\bibinfo {author} {\bibfnamefont {C.}~\bibnamefont
  {Caloz}}, \bibinfo {author} {\bibfnamefont {T.}~\bibnamefont {Itoh}}, \ and\
  \bibinfo {author} {\bibfnamefont {A.}~\bibnamefont {Rennings}},\ }\bibfield
  {title} {\enquote {\bibinfo {title} {{CRLH} metamaterial leaky-wave and
  resonant antennas},}\ }\href@noop {} {\bibfield  {journal} {\bibinfo
  {journal} {IEEE Antennas Propag. Mag.}\ }\textbf {\bibinfo {volume} {50}},\
  \bibinfo {pages} {25--39} (\bibinfo {year} {2008})}\BibitemShut {NoStop}%
\bibitem [{\citenamefont {Abielmona}\ \emph {et~al.}(2011)\citenamefont
  {Abielmona}, \citenamefont {Nguyen},\ and\ \citenamefont
  {Caloz}}]{Abielmona2011}%
  \BibitemOpen
  \bibfield  {author} {\bibinfo {author} {\bibfnamefont {S.}~\bibnamefont
  {Abielmona}}, \bibinfo {author} {\bibfnamefont {H.~V.}\ \bibnamefont
  {Nguyen}}, \ and\ \bibinfo {author} {\bibfnamefont {C.}~\bibnamefont
  {Caloz}},\ }\bibfield  {title} {\enquote {\bibinfo {title} {Analog direction
  of arrival estimation using an electronically-scanned {CRLH} leaky-wave
  antenna},}\ }\href@noop {} {\bibfield  {journal} {\bibinfo  {journal} {IEEE
  Trans. Antennas Propag.}\ }\textbf {\bibinfo {volume} {59}},\ \bibinfo
  {pages} {1408--1412} (\bibinfo {year} {2011})}\BibitemShut {NoStop}%
\bibitem [{\citenamefont {Li}\ \emph {et~al.}(2011)\citenamefont {Li},
  \citenamefont {Al\`{u}},\ and\ \citenamefont {Ling}}]{Li2011}%
  \BibitemOpen
  \bibfield  {author} {\bibinfo {author} {\bibfnamefont {Y.}~\bibnamefont
  {Li}}, \bibinfo {author} {\bibfnamefont {A.}~\bibnamefont {Al\`{u}}}, \ and\
  \bibinfo {author} {\bibfnamefont {H.}~\bibnamefont {Ling}},\ }\bibfield
  {title} {\enquote {\bibinfo {title} {Investigation of leaky-wave propagation
  and radiation in a metal cut-wire array},}\ }\href@noop {} {\bibfield
  {journal} {\bibinfo  {journal} {IEEE Trans. Antennas Propag.}\ }\textbf
  {\bibinfo {volume} {60}},\ \bibinfo {pages} {1630--1634} (\bibinfo {year}
  {2011})}\BibitemShut {NoStop}%
\bibitem [{\citenamefont {Liu}\ and\ \citenamefont {Al\`{u}}(2010)}]{Liu2010}%
  \BibitemOpen
  \bibfield  {author} {\bibinfo {author} {\bibfnamefont {X.-X.}\ \bibnamefont
  {Liu}}\ and\ \bibinfo {author} {\bibfnamefont {A.}~\bibnamefont {Al\`{u}}},\
  }\bibfield  {title} {\enquote {\bibinfo {title} {Subwavelength leaky-wave
  optical nanoantennas: directive radiation from linear arrays of plasmonic
  nanoparticles},}\ }\href@noop {} {\bibfield  {journal} {\bibinfo  {journal}
  {Phys. Rev. B}\ }\textbf {\bibinfo {volume} {82}},\ \bibinfo {pages} {144305}
  (\bibinfo {year} {2010})}\BibitemShut {NoStop}%
\bibitem [{\citenamefont {Naify}\ \emph {et~al.}(2013)\citenamefont {Naify},
  \citenamefont {Layman}, \citenamefont {Martin}, \citenamefont {Nicholas},
  \citenamefont {Calvo},\ and\ \citenamefont {Orris}}]{Naify2013}%
  \BibitemOpen
  \bibfield  {author} {\bibinfo {author} {\bibfnamefont {C.~J.}\ \bibnamefont
  {Naify}}, \bibinfo {author} {\bibfnamefont {C.~N.}\ \bibnamefont {Layman}},
  \bibinfo {author} {\bibfnamefont {T.~P.}\ \bibnamefont {Martin}}, \bibinfo
  {author} {\bibfnamefont {M.}~\bibnamefont {Nicholas}}, \bibinfo {author}
  {\bibfnamefont {D.~C.}\ \bibnamefont {Calvo}}, \ and\ \bibinfo {author}
  {\bibfnamefont {G.~J.}\ \bibnamefont {Orris}},\ }\bibfield  {title} {\enquote
  {\bibinfo {title} {Experimental realization of a variable index transmission
  line metamaterial as an acoustic leaky-wave antenna},}\ }\href@noop {}
  {\bibfield  {journal} {\bibinfo  {journal} {Appl. Phys. Lett.}\ }\textbf
  {\bibinfo {volume} {102}},\ \bibinfo {pages} {203508} (\bibinfo {year}
  {2013})}\BibitemShut {NoStop}%
\bibitem [{\citenamefont {Esfahlani}\ \emph {et~al.}(2016)\citenamefont
  {Esfahlani}, \citenamefont {Karkar}, \citenamefont {Lissek},\ and\
  \citenamefont {Mosig}}]{Esfahlani2016}%
  \BibitemOpen
  \bibfield  {author} {\bibinfo {author} {\bibfnamefont {H.}~\bibnamefont
  {Esfahlani}}, \bibinfo {author} {\bibfnamefont {S.}~\bibnamefont {Karkar}},
  \bibinfo {author} {\bibfnamefont {H.}~\bibnamefont {Lissek}}, \ and\ \bibinfo
  {author} {\bibfnamefont {J.~R.}\ \bibnamefont {Mosig}},\ }\bibfield  {title}
  {\enquote {\bibinfo {title} {Acoustic dispersive prism},}\ }\href@noop {}
  {\bibfield  {journal} {\bibinfo  {journal} {Sci. Rep.}\ }\textbf {\bibinfo
  {volume} {6}},\ \bibinfo {pages} {18911} (\bibinfo {year}
  {2016})}\BibitemShut {NoStop}%
\bibitem [{\citenamefont {Naify}\ \emph {et~al.}(2015)\citenamefont {Naify},
  \citenamefont {Guild}, \citenamefont {Rohde}, \citenamefont {Calvo},\ and\
  \citenamefont {Orris}}]{Naify2015}%
  \BibitemOpen
  \bibfield  {author} {\bibinfo {author} {\bibfnamefont {C.~J.}\ \bibnamefont
  {Naify}}, \bibinfo {author} {\bibfnamefont {M.~D.}\ \bibnamefont {Guild}},
  \bibinfo {author} {\bibfnamefont {C.~A.}\ \bibnamefont {Rohde}}, \bibinfo
  {author} {\bibfnamefont {D.~C.}\ \bibnamefont {Calvo}}, \ and\ \bibinfo
  {author} {\bibfnamefont {G.~J.}\ \bibnamefont {Orris}},\ }\bibfield  {title}
  {\enquote {\bibinfo {title} {Demonstration of a directional sonic prism in
  two dimensions using an air-acoustic leaky wave antenna},}\ }\href@noop {}
  {\bibfield  {journal} {\bibinfo  {journal} {Appl. Phys. Lett.}\ }\textbf
  {\bibinfo {volume} {107}},\ \bibinfo {pages} {133505} (\bibinfo {year}
  {2015})}\BibitemShut {NoStop}%
\bibitem [{\citenamefont {Naify}\ \emph {et~al.}(2016)\citenamefont {Naify},
  \citenamefont {Rohde}, \citenamefont {Martin}, \citenamefont {Nicholas},
  \citenamefont {Guild},\ and\ \citenamefont {Orris}}]{Naify2016}%
  \BibitemOpen
  \bibfield  {author} {\bibinfo {author} {\bibfnamefont {C.~J.}\ \bibnamefont
  {Naify}}, \bibinfo {author} {\bibfnamefont {C.~A.}\ \bibnamefont {Rohde}},
  \bibinfo {author} {\bibfnamefont {T.~P.}\ \bibnamefont {Martin}}, \bibinfo
  {author} {\bibfnamefont {M.}~\bibnamefont {Nicholas}}, \bibinfo {author}
  {\bibfnamefont {M.~D.}\ \bibnamefont {Guild}}, \ and\ \bibinfo {author}
  {\bibfnamefont {G.~J.}\ \bibnamefont {Orris}},\ }\bibfield  {title} {\enquote
  {\bibinfo {title} {Generation of topologically diverse acoustic vortex beams
  using a compact metamaterial aperture},}\ }\href@noop {} {\bibfield
  {journal} {\bibinfo  {journal} {Appl. Phys. Lett.}\ }\textbf {\bibinfo
  {volume} {108}},\ \bibinfo {pages} {223503} (\bibinfo {year}
  {2016})}\BibitemShut {NoStop}%
\bibitem [{\citenamefont {Bongard}\ \emph {et~al.}(2010)\citenamefont
  {Bongard}, \citenamefont {Lissek},\ and\ \citenamefont
  {Mosig}}]{Bongard2010}%
  \BibitemOpen
  \bibfield  {author} {\bibinfo {author} {\bibfnamefont {F.}~\bibnamefont
  {Bongard}}, \bibinfo {author} {\bibfnamefont {H.}~\bibnamefont {Lissek}}, \
  and\ \bibinfo {author} {\bibfnamefont {J.~R.}\ \bibnamefont {Mosig}},\
  }\bibfield  {title} {\enquote {\bibinfo {title} {Acoustic transmission line
  metamaterial with negative/zero/positive refractive index},}\ }\href@noop {}
  {\bibfield  {journal} {\bibinfo  {journal} {Phys. Rev. B}\ }\textbf {\bibinfo
  {volume} {82}},\ \bibinfo {pages} {094306} (\bibinfo {year}
  {2010})}\BibitemShut {NoStop}%
\bibitem [{\citenamefont {Bongard}(2009)}]{BongardThesis}%
  \BibitemOpen
  \bibfield  {author} {\bibinfo {author} {\bibfnamefont {F.}~\bibnamefont
  {Bongard}},\ }\emph {\bibinfo {title} {Contribution to characterization
  techniques for practical metamaterials and microwave applications}},\
  \href@noop {} {Ph.D. thesis},\ \bibinfo  {school} {Ecole Polytechnique
  Federale de Lausanne} (\bibinfo {year} {2009})\BibitemShut {NoStop}%
\bibitem [{\citenamefont {Morse}(1948)}]{Morse}%
  \BibitemOpen
  \bibfield  {author} {\bibinfo {author} {\bibfnamefont {P.~M.}\ \bibnamefont
  {Morse}},\ }\href@noop {} {\emph {\bibinfo {title} {Vibration and sound}}},\
  \bibinfo {edition} {2nd}\ ed.\ (\bibinfo  {publisher} {McGraw-Hill},\
  \bibinfo {address} {New York},\ \bibinfo {year} {1948})\BibitemShut {NoStop}%
\bibitem [{\citenamefont {Blackstock}(2000)}]{Blackstock}%
  \BibitemOpen
  \bibfield  {author} {\bibinfo {author} {\bibfnamefont {D.~T.}\ \bibnamefont
  {Blackstock}},\ }\href@noop {} {\emph {\bibinfo {title} {Fundamentals of
  Physical Acoustics}}},\ \bibinfo {edition} {1st}\ ed.\ (\bibinfo  {publisher}
  {John Wiley \& Sons},\ \bibinfo {address} {New York},\ \bibinfo {year}
  {2000})\BibitemShut {NoStop}%
\bibitem [{\citenamefont {Mellow}\ and\ \citenamefont
  {K\"{a}rkk\"{a}inen}(2011)}]{Mellow2011}%
  \BibitemOpen
  \bibfield  {author} {\bibinfo {author} {\bibfnamefont {T.}~\bibnamefont
  {Mellow}}\ and\ \bibinfo {author} {\bibfnamefont {L.}~\bibnamefont
  {K\"{a}rkk\"{a}inen}},\ }\bibfield  {title} {\enquote {\bibinfo {title} {On
  the sound fields of infinitely long strips},}\ }\href@noop {} {\bibfield
  {journal} {\bibinfo  {journal} {J. Acoust. Soc. Am.}\ }\textbf {\bibinfo
  {volume} {130}},\ \bibinfo {pages} {153--167} (\bibinfo {year}
  {2011})}\BibitemShut {NoStop}%
\bibitem [{\citenamefont {Ingard}(1953)}]{Ingard1953}%
  \BibitemOpen
  \bibfield  {author} {\bibinfo {author} {\bibfnamefont {U.}~\bibnamefont
  {Ingard}},\ }\bibfield  {title} {\enquote {\bibinfo {title} {On the theory
  and design of acoustic resonators},}\ }\href@noop {} {\bibfield  {journal}
  {\bibinfo  {journal} {J. Acoust. Soc. Am.}\ }\textbf {\bibinfo {volume}
  {25}},\ \bibinfo {pages} {1037--1061} (\bibinfo {year} {1953})}\BibitemShut
  {NoStop}%
\bibitem [{\citenamefont {Fahy}(2001)}]{Fahy2001}%
  \BibitemOpen
  \bibfield  {author} {\bibinfo {author} {\bibfnamefont {F.~J.}\ \bibnamefont
  {Fahy}},\ }\href@noop {} {\emph {\bibinfo {title} {Foundations of
  {E}ngineering {A}coustics}}}\ (\bibinfo  {publisher} {Academic Press},\
  \bibinfo {address} {London},\ \bibinfo {year} {2001})\BibitemShut {NoStop}%
\bibitem [{\citenamefont {Sherman}\ and\ \citenamefont
  {Butler}(2007)}]{ShermanButler}%
  \BibitemOpen
  \bibfield  {author} {\bibinfo {author} {\bibfnamefont {C.~H.}\ \bibnamefont
  {Sherman}}\ and\ \bibinfo {author} {\bibfnamefont {J.~L.}\ \bibnamefont
  {Butler}},\ }\href@noop {} {\emph {\bibinfo {title} {Transducers and arrays
  for underwater sound}}}\ (\bibinfo  {publisher} {Springer},\ \bibinfo
  {address} {New York},\ \bibinfo {year} {2007})\BibitemShut {NoStop}%
\bibitem [{\citenamefont {Mast}(2007)}]{Mast2007}%
  \BibitemOpen
  \bibfield  {author} {\bibinfo {author} {\bibfnamefont {T.~D.}\ \bibnamefont
  {Mast}},\ }\bibfield  {title} {\enquote {\bibinfo {title} {Fresnel
  approximations for acoustic fields of rectangularly symmetric sources},}\
  }\href@noop {} {\bibfield  {journal} {\bibinfo  {journal} {J. Acoust. Soc.
  Am.}\ }\textbf {\bibinfo {volume} {121}},\ \bibinfo {pages} {3311--3322}
  (\bibinfo {year} {2007})}\BibitemShut {NoStop}%
\bibitem [{\citenamefont {Kino}(1987)}]{Kino}%
  \BibitemOpen
  \bibfield  {author} {\bibinfo {author} {\bibfnamefont {G.~S.}\ \bibnamefont
  {Kino}},\ }\href@noop {} {\emph {\bibinfo {title} {Acoustic waves, devices,
  imaging and analog signal processing}}}\ (\bibinfo  {publisher}
  {Prentice-Hall},\ \bibinfo {address} {Englewood Cliffs, New Jersey},\
  \bibinfo {year} {1987})\BibitemShut {NoStop}%
\end{thebibliography}

%

\newpage
\printtables
\newpage
\printfigures
\newpage

\end{document}